\documentclass[12pt]{article} 
\makeatletter \@addtoreset{equation}{section} \makeatother
\renewcommand{\theequation}{\thesection.\arabic{equation}}
\newcommand{\ba}{\begin{array}}
\newcommand{\ea}{\end{array}}
\newcommand{\beq}{\begin{equation}}
\newcommand{\eeq}{\end{equation}}
\newcommand{\bea}{\begin{eqnarray}}
\newcommand{\eea}{\end{eqnarray}}

\def\bce{\begin{center}}
\def\ece{\end{center}}

\def\nonu{\nonumber}

\def\pa{\partial}
\def\al{\alpha}
\def\be{\beta}

\def\ep{\epsilon}

\def\la{\lambda}

\def\eps6{{\displaystyle \mathop{\epsilon}^{6}}{}}
\def\g6{{\displaystyle \mathop{g}^{6}}{}}

\def\nab6{{\displaystyle \mathop{\nabla}^{6}}{}}

\def\0{{\sst{(0)}}}
\def\1{{\sst{(1)}}}
\def\2{{\sst{(2)}}}
\def\3{{\sst{(3)}}}
\def\4{{\sst{(4)}}}
\def\5{{\sst{(5)}}}
\def\6{{\sst{(6)}}}
\def\7{{\sst{(7)}}}
\def\8{{\sst{(8)}}}

\def\ba{\begin{array}}
\def\ea{\end{array}}
\def\beq{\begin{equation}}
\def\eeq{\end{equation}}
\def\be{\begin{equation}}
\def\ee{\end{equation}}

\def\Tr{\mathop{\rm Tr}}

\def\la{\lambda}
\def\eps{\epsilon}

\def\ba{\begin{array}}
\def\ea{\end{array}}
\def\beq{\begin{equation}}
\def\eeq{\end{equation}}
\def\be{\begin{equation}}
\def\ee{\end{equation}}

\def\Tr{\mathop{\rm Tr}}

\def\la{\lambda}
\def\eps{\epsilon}

\def\eps6{{\displaystyle \mathop{\epsilon}^{6}}{}}

\def\nab6{{\displaystyle \mathop{\nabla}^{6}}{}}

\newcommand{\bean}{\begin{eqnarray*}}
\newcommand{\eean}{\end{eqnarray*}}

\begin{document}
\thispagestyle{empty} \addtocounter{page}{-1}
   \begin{flushright}
\end{flushright}

\vspace*{1.3cm}
  
\centerline{ \Large \bf
  A Supersymmetric  Enhancement of}
\vspace*{0.5cm}
\centerline{ \Large \bf
${\cal N}=1$ Holographic Minimal Model} 
\vspace*{1.5cm}
\centerline{\bf Changhyun Ahn
    and Jinsub Paeng } 
\vspace*{1.0cm} 
 \centerline{\it 
   Department of Physics, Kyungpook National University, Taegu
41566, Korea} 
 \vspace*{1.0cm} 
 \centerline{\tt ahn@knu.ac.kr,
   \qquad jdp2r@knu.ac.kr}

\vskip2cm

\centerline{\bf Abstract}
\vspace*{0.5cm}

By studying the ${\cal N}=1$ holographic minimal model at the ``critical'' level, we  obtain the lowest ${\cal N}=2$ higher spin multiplet of spins $(\frac{3}{2}, 2, 2, \frac{5}{2})$  in terms of two adjoint fermion types for generic $N$. We subsequently determine operator product expansions between the lowest and
second lowest (${\cal N}=2$) higher
spin multiplet of spins $(3, \frac{7}{2}, \frac{7}{2}, 4)$, and the corresponding Vasiliev's oscillator formalism with matrix generalization on $AdS_3$ higher spin theory in the extension of $OSp(2|2)$ superconformal algebra. Under the large $N$ limit (equivalent to large central charge) in the extension of ${\cal N}=2$ superconformal algebra in two dimensions,
operator product expansions
provide asymptotic symmetry algebra in $AdS_3$ higher spin theory.

\baselineskip=18pt
\newpage
\renewcommand{\theequation}
{\arabic{section}\mbox{.}\arabic{equation}}

\tableofcontents

\section{Introduction}

WZW primary fields are also Virasoro primary fields.
Since the Virasoro zero mode  acting on the primary state is proportional to the quadratic Casimir operator of the finite Lie algebra, primary field  conformal weight (or spin)~\cite{BS,DMS} is half the quadratic Casimir eigenvalues divided by the sum of the level and dual Coxeter number of the finite Lie algebra. When the level equals the dual Coxeter number in the adjoint representation, conformal weight is halved\footnote{ For example,  the quadratic Casimir operator eigenvalue = $2N$ and the dual Coxeter number = $N$ for $SU(N)$, where the overall numerical factor $\frac{1}{2}$ is the conformal weight (or spin) of the adjoint fermion.}. For the diagonal coset model described in~\cite{BS}, Section~$7.3$, spin-$\frac{3}{2}$ current, which is a ${\cal N}=1$ supersymmetry generator  and commutes with the diagonal spin-$1$ current, can be determined as the linear combination of two spin-$1$ current types and adjoint fermions~\cite{Douglas,BBSScon,GS88} (see, for example,~\cite{BS}, $(7.50)$). This leads to coset construction of ${\cal N}=1$ superconformal algebra~\cite{GKO} for $SU(2)$, coset construction of ${\cal N}=1$  $W_3$ algebra for $SU(3)$~\cite{HR,ASS,SS,Ahn1211}, and ${\cal N}=1$ higher spin multiplets for $SU(N)$~\cite{Ahn1305}. Although the ${\cal N}=1$ extension for bosonic coset models can be obtained from the particular level condition, we can also consider other cases with ${\cal N}=2$ and ${\cal N}=3$ extensions\footnote{ Taking adjoint spin-$\frac{1}{2}$  fermions in the second factor appearing in the numerator of the diagonal coset model~\cite{BS} provides the coset construction for ${\cal N}=2$ superconformal algebra~\cite{BFK}, and the observed ${\cal N}=2$ higher spin currents can be determined~\cite{GHKSS,Ahn1604}. Creutzig et al. and Ahn et al.~\cite{CHR1406,AK1607} investigated the ${\cal N} =3 $ extension from the ${\cal N}=2$ coset model.}.

Gaberdiel and Gopakumar~\cite{GG1011} proposed duality between higher spin gauge theory on $AdS_3$ space~\cite{PV9806,PV9812} and the large $N$ 't Hooft limit of a family of $W_N$ minimal models (see also~\cite{GG1205,GG1207,GG1406,GG1512}). This is a natural analogue of Klebanov and Polyakov duality~\cite{KP} relating the $O(N)$ vector model in 3-dimensions to a higher spin theory on $AdS_4$ space. Then we can generalize~\cite{GG1011} by considering Klebanov and Polyakov duality in one lower dimension.  The relevant coset  model was derived by replacing the $SU(N)$ group by $SO(2N)$ or $SO(2N+1)$~\cite{LF1990,GKOplb}, and ${\cal N}=1$ and ${\cal N}=2$ extensions of the (bosonic) orthogonal coset model obtained~\cite{Ahn1701} (see~\cite{KS1606} also). Thus, putting the above level condition into the ${\cal N}=1$ coset model, we can obtain  the ${\cal N}=2$ extension of the supersymmetric coset model.

Therefore, this paper considers the following coset model~\cite{CHR} at the ``critical'' level,
\bea  
\frac{G}{H} =
\frac{\widehat{SO}(2N+1)_k \oplus \widehat{SO}(2N)_1}
     {\widehat{SO}(2N)_{k+1}}\,\,\, \mbox{with $k=2N-1$}.
\label{coset}
\eea
For $SO(2N+1)$, the quadratic Casimir eigenvalue for the adjoint representation = $(2N-1)$, i.e., the dual Coxeter number of $SO(2N+1)$. The central charge of the coset model at the critical level is 
\bea
c = \frac{3Nk}{(k+2N-1)} \Bigg|_{k=2N-1}= \frac{3N}{2},
\label{centralcharge}
\eea
where the infinity limit of the central charge is equivalent to the infinity limit of $N$. Additional adjoint fermions occur in the first and second factors of group $G$ in (\ref{coset}).

We would like to construct the additional higher spin currents and their operator product expansions (OPEs) for (\ref{coset}).  We assume ``minimal'' ${\cal N}=2$ extension of the ${\cal N}=1$ higher spin currents~\cite{CHR,AP1310,CV1305},
\bea
&& ({\it 1, \frac{3}{2}}, \frac{3}{2},2);
({\it \frac{3}{2},2}, 2, {\frac{5}{2}}), ({\it 3,
\frac{7}{2}},\frac{7}{2}, 4), ({\it \frac{7}{2}, 4}, 4,
\frac{9}{2}), ({\it 5, \frac{11}{2}}, \frac{11}{2}, 6),
\cdots, \nonu \\
&&
({\it n-\frac{1}{2}, n},n,n+\frac{1}{2}),
({\it n+1,n+\frac{3}{2}}, n+\frac{3}{2},
n+2), \cdots, 
\label{fields}
\eea
where $n =2,4, 6, \cdots$. The first multiplet is the well-known ${\cal N}=2$ superconformal algebra generator~\cite{BFK}. The first two components of each ${\cal N}=2$ (higher spin) multiplet in (\ref{fields}) are new and superpartners of the last two components. Thus, we obtain   the lowest ${\cal N}=2$ higher spin multiplet of spins $(\frac{3}{2}, 2, 2, \frac{5}{2})$  in terms of two adjoint fermion types for generic $N$, their OPEs, and  OPEs between the lowest and second ${\cal N}=2$ lowest higher
spin multiplet of spins $(3, \frac{7}{2}, \frac{7}{2}, 4)$.

We also construct higher spin algebra generators in terms of oscillators corresponding to the first two higher spin multiplets in (\ref{fields}), from OPEs realized in (\ref{coset}), and explicitly provide some related (anti)commutators. The higher spin-$\frac{3}{2}$ current in (\ref{fields}) requires additional degrees of freedom because the supersymmetry generator of spin-$\frac{3}{2}$ also has spin-$\frac{3}{2}$, leaving no room for the higher spin-$\frac{3}{2}$ generator because they share a linear term in the oscillator and we cannot differentiate them. This requires matrix generalization of the Vasiliev theory~\cite{CHR1406}.

An interesting question is how higher spin symmetry (with supersymmetry) for Vasiliev higher spin theory on $AdS_3$ space  appears in string theory. The ${\cal N}=4$ extension of~\cite{GG1011} was described in~\cite{GG1305} with the hope that higher spin dualities might be embedded in the string dualities (see also~\cite{GG1406,GG1512}). Although some observations in the presence of infinite tower of modes become massless~\cite{GGH,FGJ,EGL,GHKPR,GG1803}, emergence of higher spin symmetry has not been fully clarified from string theory viewpoint. There are few examples on string theory with ${\cal N}=2$ supersymmetry compared to ${\cal N}=3$ or ${\cal N}=4$ supersymmetry. Recently, ${\cal N}=2$ supergravity solutions for $10$-dimensional theory containing $AdS_3$ space have been found~\cite{Eberhardt}, following~\cite{DEG}, and  an important supergravity outcome is that the symmetric orbifold of the (four dimensional) hyperelliptic surface supports higher spin symmetry. Thus, it remains an open problem to obtain the corresponding string theory containing ${\cal N}=2$ supergravity. Perhaps chiral primary states~\cite{GHKSS} of the present model at the critical level will be useful in this regard, which is one of the motivations for the current paper.
Datta et al.~\cite{DEG} showed that the generalized (\ref{coset}),
replacing numerical value $1$ by arbitrary $M$, can be  used to obtain
the corresponding (unknown) string theory.
We expect that the current paper outcomes will provide further direction to help construct the unknown string theory. Descriptions from~\cite{CHR1406} regarding two-dimensional conformal field theory will also be useful to understand this general coset.
The remainder of this paper is organized as follows
(the Thielemans pacakge~\cite{Thielemans} is used)
\begin{itemize}
\item Section $2$ realizes the four ${\cal N}=2$ superconformal algebra currents in terms of the two adjoint fermion types in (\ref{coset}).
\item Section $3$ analyzes the first four ${\cal N}=2$ higher spin multiplets in (\ref{fields}) for fixed $N=4$. 
\item Section $4$ determines the lowest ${\cal N}=2$ higher spin multiplet in terms of two adjoint fermions for generic $N$, and obtains its OPE and component results. The asymptotic symmetry algebra~\cite{HR-1,CFPT,GH,CFP,HLPR,HP} for the matrix extension of Vasiliev theory is also obtained by taking the large central charge limit (\ref{centralcharge}).  OPE between the lowest ${\cal N}=2$ higher spin multiplet and the second ${\cal N}=2$ higher spin multiplet is obtained using the Jacobi identity.

\item Section $5$ reviews the ``wedge'' algebra of ${\cal N}=2$ superconformal algebra, and constructs   $OSp(2|2)$ higher spin algebra generators in terms of oscillators and their algebra~\cite{PV9806,PV9812,BVdplb,BVdnpb}. We also analyze the match between these outcomes and  Section $4$ under the large $c$ limit with wedge condition.

\item Section $6$ summarizes the findings and concludes the paper, and briefly discusses some remaining open problems.

\item Appendices $A$--$F$ present various details described in the previous sections.

\end{itemize}

\section{Four currents for ${\cal N}=2$ superconformal algebra  }

We construct the four currents for ${\cal N}=2$ superconformal algebra in the coset model (\ref{coset}) following~\cite{Ahn1604}.

\subsection{Kac-Moody spin-$1$ currents}

For the diagonal coset model in (\ref{coset}), the spin-$1$ current $J^A(z)$ with level $k$ and spin-$\frac{1}{2}$ current $\chi^i(z)$ (whose spin-$1$ current has  the level $1$) generate the affine Lie algebra $G=\widehat{SO}(2N+1)_{k} \oplus \widehat{SO}(2N)_1$~\cite{CHR}.
Adjoint indices $A, B,  \cdots$ of $SO(2N+1)$ = $1, 2, \cdots, N(2N+1)$ and vector indices $i, j, \cdots $ of $SO(2N)$ = $1, 2, \cdots, 2N$ and can be relabeled by adding $N(2N-1)$ respectively\footnote{Adjoint indices $A, B, \cdots$ of $SO(2N+1)$ can be further decomposed into $SO(2N)$ adjoint indices  $a, b, \cdots = 1, 2, \cdots, N(2N-1)$ and $SO(2N)$ vector indices $i, j, \cdots =1+N(2N-1), \cdots, 2N+N(2N-1)=N(2N+1)$, i.e.,  $SO(2N+1)$ adjoint indices $ A, B, \cdots =1, 2, \cdots, N(2N-1), 1+N(2N-1), \cdots, 2N+N(2N-1)$.}. Diagonal spin-$1$ current $(J^a +K^a)(z)$ with level $(k+1)$, where $K^a(z)$ is quadratic in spin-$\frac{1}{2}$ current (\ref{spin1currents}), generates $H=\widehat{SO}(2N)_{k+1}$ affine Lie algebra.

From condition $k=2N-1$, we can introduce fermions $\psi^A(z)$ in the first factor of $G$, and then  consider two fermion field types $\psi^A(z)$ and $\chi^i(z) $ that satisfy the OPEs
\bea
\psi^A(z) \, \psi^B(w) & = & -\frac{1}{(z-w)} \,
\frac{1}{2} \, \delta^{AB} + \cdots, \qquad A \equiv (a, i),
\nonu \\
\chi^i(z) \, \chi^j(w) & = & - \frac{1}{(z-w)} \,
\frac{1}{2} \, \delta^{ij} + \cdots.
\label{psichi}
\eea
The corresponding  Kac-Moody spin-1 currents are 
\bea
J^a(z)  & \equiv &  f^{aBC} \, \psi^B  \psi^C(z)=
 f^{abc} \, \psi^b  \psi^c(z) +  f^{aij} \, \psi^i  \psi^j(z),
\nonu \\
J^i(z)  & \equiv &  f^{iBC} \, \psi^B  \psi^C(z)=
2 f^{ija} \, \psi^j  \psi^a(z),
\nonu \\
K^a(z) & \equiv &  f^{aij} \, \chi^i  \chi^j(z),
\label{spin1currents}
\eea
hence $ f^{ijk} =0= f^{iab}$, which obey the nontrivial OPEs from (\ref{psichi}) and (\ref{spin1currents}),
\bea
J^a(z) \, J^b(w) & = & - \frac{1}{(z-w)^2} \, (2N-1) \, \delta^{ab}
+ \frac{1}{(z-w)} \, f^{abc} \, J^c(w) + \cdots,
\nonu \\
J^a(z) \, J^i(w) & = &  \frac{1}{(z-w)} \, f^{aij} \, J^j(w) + \cdots,
\nonu \\
J^i(z) \, J^j(w) & = & - \frac{1}{(z-w)^2} \, (2N-1) \, \delta^{ij}
+ \frac{1}{(z-w)} \, f^{ija} \, J^a(w) + \cdots,
\nonu \\
K^a(z) \, K^b(w) & = & -\frac{1}{(z-w)^2} \, \delta^{ab}
+ \frac{1}{(z-w)} \, f^{abc} \, K^c(w) + \cdots.
\label{JKOPE}
\eea
Spin-$1$ current $J^A(z)$ has level $(2N-1)$. Hence, adding the first and last in (\ref{JKOPE}), diagonal spin-$1$ current $(J^a+K^a)(z)$ has level $2N$.

\subsection{Four currents in terms of fermions}

This section presents the four ${\cal N}=2$ superconformal algebra currents.

$\bullet$ Coset spin-$1$ current

Using the $SO(2N)$ invariant tensor of rank $2$, spin-$1$ current can be expressed as 
\bea
J(z) =
  i\, \delta^{ij}\, \psi^i  \chi^j(z)
=
i\, \psi^i  \chi^i(z).
\label{t1}
\eea
Thus, the OPE between spin-$1$ current and itself is 
\bea
J(z) \, J(w) &=& \frac{1}{(z-w)^2} \, \frac{c}{3} + \cdots,
\qquad
c = \frac{3}{2} \, N.
\label{JJOPE}
\eea
The overall factor in (\ref{t1}) is fixed by requiring the central term of OPE $J(z) \, J(w)$ should behave as in (\ref{JJOPE}). The coset spin-$1$ current has no singular terms with the diagonal spin-$1$ current, i.e., 
\bea
(J^a+K^a)(z) \, J(w) & = &  + \cdots.
\label{regularcondition}
\eea
In particular, combining the two fermions has nonzero $U(1)$ charge, $\pm \frac{1}{2}$, associated with the spin-$1$ current
\bea
J(z) \, (\psi^i  \pm i \chi^i)(w) & = &  \pm \frac{1}{(z-w)} \,
\frac{1}{2} (\psi^i  \pm i \chi^i)(w)
+\cdots.
\label{qcharges}
\eea
Similarly, the regular term in the OPE between spin-$1$ current $J(z)$ and fermion $\psi^a(w)$ can be expressed as 
\bea
J(z) \, \psi^a(w) & = & 
+\cdots.
\label{psiacharges}
\eea
We can analyze higher spin currents for fixed $U(1)$ charges in terms of fermions using fermion $U(1)$ charges in (\ref{qcharges}) and (\ref{psiacharges}).

$\bullet$ Coset spin-$\frac{3}{2}$ currents

From (\ref{regularcondition}), spin-$\frac{3}{2}$ currents should  satisfy
\bea
(J^a+K^a)(z) \, G^{\pm}(w) & = & +\cdots ,
\label{reg}
\eea
where the OPE with spin-$1$ current is 
\bea
J(z) \,  G^{\pm}(w) & = & \pm\,\frac{1}{(z-w)} \,  G^{\pm}(w)
+\cdots.
\label{JGOPE}
\eea
Then the spin-$\frac{3}{2}$ currents  of $U(1)$ charges $\pm 1$ (\ref{JGOPE}) can be expressed as 
\bea
G^{\pm}(z) =
\frac{1}{4\sqrt{2N-1}}
\Biggr[ \mp i \,\psi^i  J^i  \pm 2i \, \psi^a  K^a +  2 \chi^i  J^i
 \Biggr](z),
\label{deetee}
\eea
and we can obtain ${\cal N}=1$ spin-$\frac{3}{2}$ current by adding these two spin-$\frac{3}{2}$ currents.



$\bullet$ Coset spin-$2$ current

The spin-$2$ stress energy tensor that satisfies the regular condition with the coset spin-$1$ current as in (\ref{regularcondition}) and (\ref{reg}) can be obtained from the difference between those in group $G$ and the one in subgroup $H$ with the correct coefficients,
\bea
T(z) & = & -\frac{1}{4(2N-1)} \, (J^a  +J^i J^i)(z)
-\frac{1}{2(2N-1)} \, K^a  K^a(z)
\nonu \\
& + & \frac{1}{4(2N-1)}  (J^a +K^a) (J^a+K^a)(z),
\label{T}
\eea
where the central charge is given by (\ref{JJOPE}).


Therefore, the four ${\cal N}=2$ superconformal algebra currents in the coset model are summarized by (\ref{t1}), (\ref{deetee}), and (\ref{T}). In ${\cal N}=2$ superspace, they can be organized by a single ${\cal N}=2$ stress energy tensor,
\bea
{\bf T}(Z) = J(z) + \theta \, G^{+}(z) + \bar{\theta} \, G^{-}(z) +
\theta \, \bar{\theta} \, T(z).
\label{stressN2}
\eea
The defining OPEs between the four currents in ($\ref{stressN2}$) are given by (\ref{n2scaexpression}).

\section{ ${\cal N}=2$ higher spin  currents for fixed $N=4$}
This section describes ${\cal N}=2$ higher spin multiplets in the coset model for $N=4$.  We show higher spin-$\frac{3}{2}$ (primary) current that belongs to the lowest ${\cal N}=2$ higher spin multiplet, and the presence of other higher spin currents in other ${\cal N}=2$ higher spin multiplets.

\subsection{The first (lowest) ${  \cal N}=2$ higher spin multiplet}

Creutzig et al., Ahn et al. and Candu et al.~\cite{CHR,AP1310,CV1305} showed that the lowest ${\cal N}=1$ higher spin multiplet contains higher spin-$2$ and higher spin-$\frac{5}{2}$ currents. Due to the presence of additional fermions $\psi^a(z)$ and $\psi^i(z)$, there is the possibility to have additional lower ${   \cal N}=1$ higher spin multiplet of spins $\frac{3}{2}$ and $2$, which is a superpartner of the ${   \cal N}=1$ higher spin multiplet. Thus, it is natural to check whether a higher spin-$\frac{3}{2}$ current occurs in the coset model.

Let us consider the most general spin-$\frac{3}{2}$ current with unknown $U(1)$ charge $q$ and unknown coefficients~\cite{AK1607},
\bea W_{q}^{(\frac{3}{2})}(z) = 
C_{1}^{ABC} \psi^{A}\psi^{B}\psi^{C}
+C_{2}^{iBC} \chi^{i}\psi^{B}\psi^{C}
+C_{3}^{Aij} \psi^{A} \chi^{i}\chi^{j}
+C_{4}^{ijk} \chi^{i} \chi^{j}\chi^{k}
+C_{5}^{A} \pa \psi^{A}
+C_{6}^{i} \pa \chi^{i},
\nonu 
\\  
\label{GW3}
\eea 
where the undetermined coefficients can be $SO(2N+1)$ or $SO(2N)$ (of $G$ in the coset model) invariant tensors. To find this higher spin-$\frac{3}{2}$ current (\ref{GW3}), we use the conditions from (\ref{TWq}) with the regular condition in the OPE for diagonal spin-$1$ current (see~\cite{Ahn1111,BS} for the GKO~\cite{GKO} coset construction), 
\bea
T(z) \, W_q^{(\frac{3}{2})}(w) \Bigg|_{\frac{1}{(z-w)^2}} &=& \frac{3}{2} W_q^{(\frac{3}{2})}(w),
\nonu \\
T(z) \, W_q^{(\frac{3}{2})}(w) \Bigg|_{\frac{1}{(z-w)}} &=& \pa W_q^{(\frac{3}{2})}(w),
\nonu \\
J(z) \,  W_q^{(\frac{3}{2})}(w) \Bigg|_{\frac{1}{(z-w)}} &=& q W_q^{(\frac{3}{2})}(w),
\nonu \\
(J^a+K^a)(z) \, W_q^{(\frac{3}{2})}(w) & = &  + \cdots. 
\label{condition1}
\eea
The first two conditions come from the primary field under the stress energy tensor, the third is $U(1)$ charge $q$ under spin-$1$ current, and finally the regular condition with diagonal spin-$1$ current (see the defining OPEs in (\ref{TWq})).

Only one  higher spin-$\frac{3}{2}$ primary field with $U(1)$ charge $q=0$ exists, denoted  by $W_{q=0}^{(h=\frac{3}{2})}(z)$, with explicit expression for fixed $N=4$, 
\bea
W_{0}^{(\frac{3}{2})}(z) =-\frac{5}{18}
\Biggr[ \psi^i  J^i+\frac{9}{5} \, \psi^a  K^a -
\frac{1}{5} \psi^a  J^a
 \Biggr](z),
\label{W3h}
\eea 
using previous relationships from (\ref{spin1currents}). The last term of (\ref{W3h}) does not appear in (\ref{deetee}). The general expression for generic $N$ will be considered in
Section $4.1$.

To find the other primary currents that belong to the same ${   \cal N}=2$ higher spin multiplet as $W_{0}^{(\frac{3}{2})}(w)$, we can use the defining relations in (\ref{TWq}). Given the lowest higher spin-$\frac{3}{2}$ current (\ref{W3h}), we can determine higher spin-$2$ currents of $U(1)$ charge $\pm 1$ using $G^{\pm}(z)$ in (\ref{deetee}),
\bea W_{\pm}^{(2)}(z) &=&
-\frac{ 1}{12\sqrt{7}} \Biggr[
 -\frac{3i}{2} \,J^iJ^i
 +\frac{i}{4} \, J^a  J^a
-\frac{7i}{2} \, J^a  K^a
\nonu \\
&& + \frac{7i}{4}K^a  K^a
 \mp 7\, J^a L^a
\pm 13 \, K^a  L^a
+ 7\, J J
\mp 70i\, \pa J
 \Biggr](z),
\label{W2pm}
\eea
where spin-$1$ current $L^a(z)$ considering the product of two fermions is 
\bea 
L^a(z) & \equiv &  f^{aij} \, \psi^i \chi^j(z),
\label{Lafield}
\eea
and (\ref{spin1currents}) and (\ref{t1}) are used.  The last four terms in (\ref{W2pm}) do not appear in (\ref{T}).

Similarly, the last component higher spin-$\frac{5}{2}$ current of vanishing $U(1)$ charge can be determined by the OPE between spin-$\frac{3}{2}$ and higher spin-$2$ currents,
\bea 
W_{0}^{(\frac{5}{2})}(z) &=&-\frac{1}{12}\Biggr[
2i\, f^{abc} \psi^a K^b L^c 
-3\, f^{abc} \psi^a
J^b L^c 
+6\, f^{abc} \psi^a \psi^b \psi^c J
\nonu \\
&-&6\,
\psi^a J^a J
-i\, d^{ajkb} \psi^a \psi^j \chi^k J^b
+i\, d^{ibjc} \chi^i \psi^b \psi^j K^c \Biggr](z),
\label{W5h}
\eea
where the last two terms contain the totally symmetric $d$ tensor of $SO(2N+1)$,
\bea
d^{ABCD}&=& \frac{1}{2}\Tr[
T^D T^A  T^B T^C
+T^D T^A  T^C T^B
+T^D T^B  T^A T^C 
+T^D T^B  T^C T^A
\nonu \\
&+&T^D T^C  T^A T^B
+T^D T^C  T^B T^A], \qquad A=(a,i).
\label{dtensor}
\eea
Since the $SO(2N)$ adjoint and vector indexes inside $SO(2N+1)$ range independently, there are nontrivial contributions in the last two terms of (\ref{W5h}), and we used  (\ref{spin1currents}), (\ref{t1}), and (\ref{Lafield}).
%

The three currents (\ref{W2pm}) and (\ref{W5h}) satisfy regularity conditions with diagonal spin-$1$ current,
\bea
(J^a+K^a)(z) \,W_{\pm}^{(2)}(w) & = & +\cdots,
\nonu \\
(J^a+K^a)(z) \,W_{0}^{(\frac{5}{2})}(w) & = & +\cdots.
\label{regg}
\eea

Therefore,  the four currents (\ref{W3h}), (\ref{W2pm}) and (\ref{W5h}) that satisfy (\ref{condition1}) and (\ref{regg}), are components of   the lowest ${\cal N}=2$ higher spin multiplet, 
\bea
{\bf W}_{0}^{(\frac{3}{2})}
&\equiv& \left(W_{0}^{(\frac{3}{2})}, \, W_{+}^{(2)}, \,
W_{-}^{(2)}, \,
W_{0}^{(\frac{5}{2})} \right).
\label{W3half}
\eea
We also checked the defining OPEs in (\ref{TWq}) for $h=\frac{3}{2}$ and $q=0$.

The existence of ${\bf W}_{0}^{(\frac{3}{2})}(Z)$~\cite{AP1310} strongly suggests there would be ${\cal N}=2$ higher spin multiplets in addition to the lowest ${\cal N}=2$ higher spin multiplet ${\bf W}_{0}^{(\frac{3}{2})}(Z)$ (\ref{W3half})
\bea
{\bf W}_{0}^{(3)}
&\equiv& \left(W_{0}^{(3)}, \, W_{+}^{(\frac{7}{2})}, \,
W_{-}^{(\frac{7}{2})}, \,
W_{0}^{(4)} \right),
\nonu \\
{\bf W}_{0}^{(\frac{7}{2})}
&\equiv& \left(W_{0}^{(\frac{7}{2})}, \, W_{+}^{(4)}, \,
W_{-}^{(4)}, \,
W_{0}^{(\frac{9}{2})} \right),
\nonu \\
{\bf W}_{0}^{(5)}
&\equiv& \left(W_{0}^{(5)}, \, W_{+}^{(\frac{11}{2})}, \,
W_{-}^{(\frac{11}{2})}, \,
W_{0}^{(6)} \right),\,\,\, \cdots\,\,\,.
\label{threeW}
\eea
The first multiplet of (\ref{threeW}) is the ${\cal N}=2$ extension of ${\cal N}=1$ higher spin multiplet of spins $\frac{7}{2}$ and $4$~\cite{AP1310}. The extra higher spin currents of spins $\frac{7}{2}$ and $4$ appear in the second multiplet of (\ref{threeW}). Similarly, the first two components in the third multiplet of (\ref{threeW}) are new additions.  The following sections observe the presence of some higher spin currents for fixed $N=4$.

\subsection{Second ${  \cal N}=2$ higher spin multiplet}

To investigate the existence of the second ${\cal N}=2$ higher spin multiplet ${\bf W}_{0}^{(3)}(Z)$ in (\ref{threeW}), we should show there is no other primary higher spin-$3$ current besides $W_{0}^{(3)}(z)$ in (\ref{coset}).  Therefore, we construct the most general higher spin-$3$ field $W^{(h=3)}_q(z)$ with unknown $U(1)$ charge $q$ and use conditions (\ref{condition1}) for this higher spin current as we did for $W^{(\frac{3}{2})}_{0}(z)$. Only one higher spin-$3$ (primary) current with $U(1)$ charge $q=0$ exists: the lowest higher spin current $W_{0}^{(3)}(z)$ of the second ${\cal N}=2$ higher spin multiplet, which can be expressed for fixed $N=4$ as
\bea
&&W_{0}^{(3)}(z) = \Biggr[
\frac{4}{847}\,(\psi^a K^a)( \psi^b L^b) +\frac{8}{847}\,(\psi^a
L^a)( \psi^i J^i) +\frac{20}{7623}\,(\psi^i J^i)( \chi^j J^j)
\nonu \\
&& -\frac{799}{45738} J^a (d^{ajkb} \psi^j \chi^k K^b) +\frac{3995}{320166}\, K^a (d^{ajkb}
\psi^j \chi^k K^b) +\frac{31}{6534}\, L^a (d^{abjk} \psi^b \psi^j J^k)
\nonu \\
&&
+\frac{1}{1089} J^{i} (d^{ibjc} \psi^b \chi^j J^c)
 +\frac{2327}{45738}i\,JJJ + \frac{19}{91476}i\, J^a J^a J -
\frac{17387}{640332}i\, K^a K^a J + \frac{1949}{45738}i\, L^a L^a J
\nonu \\ 
&& 
+\frac{106}{3267}i\, J^a K^a J
-\frac{157}{45738}\, J^a \pa L^a -\frac{17}{7623}\, L^a \pa J^a
-\frac{6947}{320166}\, K^a \pa L^a +\frac{646}{22869}i\, \pa^2 J \Biggr](z),
\label{W3}
\eea
using (\ref{spin1currents}), (\ref{t1}), and (\ref{Lafield}). In this case, an additional $d$ tensor (\ref{dtensor}) with two adjoint indices and two vector indices appears in the second and third lines.

The other components of ${\cal N}=2$ higher spin multiplet ${\bf W}_{0}^{(3)}(Z)$  in (\ref{threeW}) can be obtained from (\ref{TWq}), in principle, or they can  appear in the OPEs between the ${\cal N}=2$ higher spin multiplet ${\bf W}_{0}^{(\frac{3}{2})}(Z)$ components in (\ref{W3half}). For example, the higher spin-$3$ current $W_{0}^{(3)}(z)$ in (\ref{W3}) appears in the first-order pole for the OPE $W_{0}^{(\frac{5}{2})}(z) \, W_{0}^{(\frac{3}{2})}(w)$,
\bea
W_{0}^{(\frac{5}{2})}(z) \,
W_{0}^{(\frac{3}{2})}(w) & = & \frac{1}{(z-w)^3} \,
\frac{35}{6} J(w) +
\frac{1}{(z-w)^2} \, \frac{35}{6} \pa J(w)
\nonu \\
&+& \frac{1}{(z-w)} \, \Bigg[-\frac{2541i}{48}W_{0}^{(3)}+\frac{35}{12}G^{+}G^{-}
\nonu \\
&+&\frac{49}{24}JT-\frac{91}{144}JJJ -\frac{35}{24}\pa T
+\frac{175}{72} \pa^2 J \Bigg](w)+\cdots. 
\label{W5hW3h}
\eea
We can rearrange the first order pole in (\ref{W5hW3h}) in terms of various quasi primary fields.  Similarly, other component higher spin-$\frac{7}{2}$ currents $W_{\pm}^{(\frac{7}{2})}(z)$ appear in the first-order pole for OPE $W_{0}^{(\frac{5}{2})}(z) \, W_{\pm}^{(2)}(w)$ 
\bea
W_{0}^{(\frac{5}{2})}(z) \,
W_{\pm}^{(2)}(w) & = & -\frac{1}{(z-w)^3} \,
\frac{35}{3} G^{+}(w) + \frac{1}{(z-w)^2} \,
\Bigg[\mp\frac{7}{12}JG^{+}-\frac{91}{12} \pa G^{+} \Bigg](w)
\nonu \\
&+& \frac{1}{(z-w)} \, \Bigg[\frac{2541i}{48}
  W_{\pm}^{(\frac{7}{2})} -\frac{119}{24}G^{+}T
\mp\frac{35}{24}G^{+} \pa J +\frac{91}{48}JJG^{+} \mp\frac{7}{8}J \pa
G^{+}
\nonu \\
&-&\frac{91}{32} \pa^2 G^{+}\Bigg](w)+\cdots, 
\label{W5hW2PM}
\eea
which can be rearranged in terms of quasi primary fields in the second and first order poles. 

Finally, the last component higher spin-$4$ current $W_{0}^{(4)}(z)$ appears in the first-order pole for OPE $W_{0}^{(\frac{5}{2})}(z) \, W_{0}^{(\frac{5}{2})}(w)$,
\bea
W_{0}^{(\frac{5}{2})}(z) \,
W_{0}^{(\frac{5}{2})}(w)  & = & \frac{1}{(z-w)^5} \,
\frac{70}{3} + \frac{1}{(z-w)^3} \Bigg[\frac{119}{6}T-
\frac{7}{12}JJ \Bigg](w)
\nonu \\
&+& \frac{1}{(z-w)^2}\, \Bigg[-\frac{7}{12}J \pa J+\frac{119}{12} \pa
T \Bigg](w)
\nonu \\
&+&\frac{1}{(z-w)} \, \Bigg[-\frac{2541i}{48}W_{0}^{(4)}
-\frac{7}{12}G^{-} \pa G^{+} -\frac{7}{12}G^{+} \pa G^{-}
+\frac{91}{24}JG^{+} G^{-}
\nonu \\
&-& \frac{91}{48}JJT -\frac{91}{48}J \pa T -\frac{7}{16}J \pa^2 J
+\frac{119}{24}TT -\frac{35}{48}\pa J \pa J
\nonu \\
&+& \frac{91}{48} \pa^2 T \Bigg](w)+\cdots.
\label{W5hW5h}
\eea
Although the left hand side currents are equal, the occurrence of higher spin-$4$ current in the first order pole is expected, in contrast to the two bosonic currents (i.e., the higher spin-$5$ current does not arise in the OPE of the higher spin-$3$ current and itself), since, they are fermionic (from (\ref{W5hW5h})). OPEs in (\ref{W5hW2PM}) and (\ref{W5hW5h}) exhibit similar behavior in the ${\cal N}=1$ coset model~\cite{AP1310}.

Thus, (\ref{W5hW3h}), (\ref{W5hW2PM}), and (\ref{W5hW5h}) prove the existence of  second ${\cal N}=2$ higher spin  multiplet ${\bf W}_{0}^{(3)}(Z)$ from the OPE between the first ${\cal N}=2$ higher spin  multiplet ${\bf W}_{0}^{(\frac{3}{2})}(Z)$ and itself, although the analysis is incomplete. Section $4$ investigates their relationship from the Jacobi identity. 

\subsection{Third and fourth ${  \cal N}=2$ higher spin multiplets}

To show the existence of the third ${\cal N}=2$ higher spin multiplet ${\bf W}_{0}^{(\frac{7}{2})}(Z)$ in (\ref{threeW}), we must first prove the presence  of the first component of multiplet $W_{0}^{(\frac{7}{2})}(z)$. We can check that the higher spin-$\frac{7}{2}$ current $W_{0}^{(\frac{7}{2})}(z)$ appears in the second-order pole of the OPE (from the OPE between the first and second higher spin multiplets),
\bea
W_{0}^{(\frac{5}{2})}(z) \, W_{0}^{(3)}(w)  &
= &
\frac{1}{(z-w)^4} \frac{225i}{22} W_{0}^{(\frac{3}{2})}(w)
+ \frac{1}{(z-w)^3} \frac{75i}{22} \pa
W_{0}^{(\frac{3}{2})}(w)
\nonu \\
&+& \frac{1}{(z-w)^2}\, \Bigg[
  W_{0}^{(\frac{7}{2})}-\frac{32i}{33}G^{-}W_{+}^{(2)}
+\frac{32i}{33}G^{+}W_{-}^{(2)} +\frac{103i}{33}J
W_{0}^{(\frac{5}{2})} 
\nonu \\
&-&
\frac{85i}{22}J J
W_{0}^{(\frac{3}{2})}+\frac{51i}{11} W_{0}^{(\frac{3}{2})}T +\frac{i}{33}\pa^2
W_{0}^{(\frac{3}{2})}
\Bigg](w)
\nonu \\
&+&\frac{1}{(z-w)}\, \Bigg[\frac{7}{3} \pa W_{0}^{(\frac{7}{2})}
+\frac{6i}{11}G^{-} \pa W_{+}^{(2)}
-\frac{6i}{11}G^{+} \pa W_{-}^{(2)} +\frac{8i}{33}J
G^{-}W_{+}^{(2)}
\nonu \\
&-&\frac{47i}{66} J J \pa W_{0}^{(\frac{3}{2})}
-\frac{52i}{11}J  W_{0}^{(\frac{3}{2})} \pa J
 +\frac{3i}{11} J \pa W_{0}^{(\frac{5}{2})}
  +\frac{41i}{33} T \pa W_{0}^{(\frac{3}{2})}
\nonu \\
&+&  \frac{56i}{33} W_{-}^{(2)}\pa  G^{+} -\frac{56i}{33}
W_{+}^{(2)}\pa  G^{-}
 +\frac{28i}{11}  W_{0}^{(\frac{3}{2})} \pa T
 +\frac{116}{33} \pa J W_{0}^{(\frac{5}{2})}
 \nonu \\
&+& \frac{8i}{33}J G^{+}W_{-}^{(2)} -\frac{5i}{4}
\pa^3 W_{0}^{(\frac{3}{2})} \Bigg](w) + \cdots, \label{W5hW3}
\eea
where each singular term can be rearranged
in terms of various quasi primary fields as discussed above. The other components of multiplet ${\bf W}_{0}^{(\frac{7}{2})}(Z)$ can be obtained   from the defining OPEs in (\ref{TWq}) with explicit expressions for the higher spin-$\frac{7}{2}$ current $W_0^{\frac{7}{2}}(z)$ and spin-$\frac{3}{2}$ currents $G^{\pm}(w)$.


To prove the fourth ${\cal N}=2$ higher spin multiplet ${\bf W}_{0}^{(5)}(Z)$ existence in (\ref{threeW}), we must first prove the existence of the first component of multiplet $W_{0}^{(5)}(z)$. This higher spin-$5$ current $W_{0}^{(5)}(z)$ appears in the first-order pole of the OPE (from the OPE between the first and third higher spin multiplets),
\bea
W_{0}^{(\frac{5}{2})}(z) \,
W_{0}^{(\frac{7}{2})}(w)  &=& \frac{1}{(z-w)^3}\,
\frac{196}{3} W_{0}^{(3)}(w)
\nonu \\
& + & \frac{1}{(z-w)^2}\, \Bigg[
  \frac{980i }{99}G^{-} \pa G^{+}-\frac{980i
}{99}G^{+} \pa G^{-} +\frac{637i}{99}J J \pa J
\nonu \\
&-&\frac{1274i }{99}J \pa T +\frac{490i }{99}\pa J T
+\frac{112i}{11 }W_{0}^{(\frac{3}{2})}
W_{0}^{(\frac{5}{2})}
\label{W5hW7h}
\\
&+&  \frac{49 }{3} \pa W_{0}^{(3)} +
\frac{245i }{99}\pa^3
J \Bigg](w) +\frac{1}{(z-w)}\, \Bigg[
  W_{0}^{(5)}+ \cdots
  \Bigg](w)
+ \cdots.
\nonu
\eea
However, due to the complexity of the first-order pole
calculation~(\ref{W5hW7h}),
we could not  find any explicit structure for
the first-order pole\footnote{Since spin = $5$, the order of   higher spin-$5$ current is similar to $10$-th order   in the two fermion types, and there are more than one million terms   even for $N=4$ case. }. On the other hand, the first-order pole of this OPE cannot be expressed in terms of descendant or known composite fields, which suggests the existence of higher spin-$5$ current $W_{0}^{(5)}(z)$ belonging to the fourth higher spin multiplet. Therefore, we postulate the existence of fourth higher spin  multiplet  ${\bf W}_{0}^{(5)}(Z)$ in (\ref{threeW}). Section $4$
considers more algebraic structures for the $N$.

  \section{ Operator product expansions between  ${\cal N}=2$    higher spin  multiplets}

Section $3$ established the presence of ${\cal N}=2$ higher spin  multiplets for $N=4$, 
\bea
{\bf W}_{0}^{(\frac{3}{2})}
&\equiv& \left(W_{0}^{(\frac{3}{2})}, \, W_{+}^{(2)}, \,
W_{-}^{(2)}, \,
W_{0}^{(\frac{5}{2})} \right),
\nonu \\
{\bf W}_{0}^{(3)}
&\equiv& \left(W_{0}^{(3)}, \, W_{+}^{(\frac{7}{2})}, \,
W_{-}^{(\frac{7}{2})}, \,
W_{0}^{(4)} \right),
\nonu \\
{\bf W}_{0}^{(\frac{7}{2})}
&\equiv& \left(W_{0}^{(\frac{7}{2})}, \, W_{+}^{(4)}, \,
W_{-}^{(4)}, \,
W_{0}^{(\frac{9}{2})} \right),
\nonu \\
{\bf W}_{0}^{(5)}
&\equiv& \left(W_{0}^{(5)}, \, W_{+}^{(\frac{11}{2})}, \,
W_{-}^{(\frac{11}{2})}, \,
W_{0}^{(6)} \right),\,\,\, \cdots\,\,\,
\label{supermultiplets}.
\eea
Assuming that the super multiplets in (\ref{supermultiplets}) exist for $N > 4$, we now construct the (super) OPEs between the ${\cal N}=2$ higher spin multiplets in (\ref{supermultiplets}) for $N$.

\subsection{First ${\cal N}=2$ higher spin multiplet for generic $N$}


To find (super) OPE ${\bf W}^{(\frac{3}{2})}_{0}(Z_1){\bf W}^{(\frac{3}{2})}_{0}(Z_2) $, we must first find the four component fields of ${\bf W}^{(\frac{3}{2})}_{0}(Z) $ for $N= 4, 5, 6,$ and $7$ (or more $N > 7$), and extract the general form for the component field.

We can obtain the general expression of $W_{0}^{(\frac{3}{2})}(z)$ for generic $N$ from the first component field $W_{0}^{(\frac{3}{2})}(z)$ of the lowest ${\cal N}=2$ higher spin multiplet  ${\bf W}_{0}^{(\frac{3}{2})}(Z)$ for $N=4,5, 6$, and $7$, 
\bea 
W_{0}^{(\frac{3}{2})}(z) = 
 \sqrt{\frac{(3 N-2)}{12 (N-1) (2 N-1)}}
 \Biggr[ \psi^i  J^i+\frac{6(N-1)}{(3N-2)} \,
 \psi^a  K^a - \frac{2}{(3N-2)} \psi^a  J^a 
 \Biggr](z), 
\label{GW3h} 
\eea  
which generalizes (\ref{W3h}), and the order of $N$ in the denominator inside the square root is quadratic.  Thus, the four $N$ values completely determine this $N$ dependence and the numerator and denominator of relative coefficients behave linearly. Therefore, we can determine $N$ dependence\footnote{In particular, the relative coefficients of (\ref{GW3h}) for $N=4,5,6,7$   can be found from the last two relations in (\ref{condition1}) to be fractional   functions of   $N$, and $N$ dependence appears in the   second order pole of the OPE between $(J^a+K^a)(z)$   and $(\psi^a  J^a)(w)$ linearly. Thus, the numerator and   denominator of the fractional functions behave as $N$.   

We need four different $N$ values because the numerator goes as $c_1 \, N + c_2$ and the denominator   as $c_3 \, N +c_4$. Therefore, four distinct $N$ values   fix the unknown values $c_1,c_2,c_3$, and $c_4$. The   overall factor is fixed by requiring that the third order pole   of the higher spin-$\frac{3}{2}$ current and itself should be   $\frac{2c}{3}$, which behaves as $N$. Finally, the overall factor   can be fixed from the four values of $N$.

  We can determine the relative coefficients and overall factor by calculating the corresponding OPEs~\cite{Ahn1604}. The third order pole of the OPE between the first and the last terms is $N(2N-1)$ and the corresponding value for the OPE between the second and itself is $\frac{1}{2}\, N(2N-1)$, where the last term is $\frac{1}{2} \, N (2 N-1) (6 N-5)$. Finally, the first term is $2N(2N-1)$. There are also two relations $-2A_1-A_2+A_3(-6N+5)=0$ and $A_1 -\frac{1}{2} \, A_2 +\frac{1}{2}\, A_3 =0$, where $A_i$ are the coefficients of the higher spin-$\frac{3}{2}$ current from the last two equations of (\ref{condition1}). Then we obtain the three coefficients $A_i$ as in (\ref{GW3h}). We can proceed for the other higher spin currents similarly.}.

The other component fields of  ${\bf W}_{0}^{(\frac{3}{2})}(Z)$ for $N$ are obtained from (\ref{TWq}) with $G^{\pm}$,
\bea 
W_{\pm}^{(2)}(z) &=&
\frac{\sqrt{3}}{2(2N-1)\sqrt{(N-1)(3N-2)}}
\Biggr[
 -\frac{i(N-1)}{2} \,J^iJ^i
 +\frac{i}{4} \, J^a  J^a
-\frac{(2N-1)i}{2} \, J^a  K^a
\nonu \\
&& + \frac{(2N-1)i}{4}K^a  K^a
 \mp (2N-1)\, J^a L^a
\pm (4N-3) \, K^a  L^a
\nonu \\
&&
 + (2N-1)\, J J
\mp i(2N-1)(3N-2)\, \pa J
 \Biggr](z),
\nonu \\
W_{0}^{(\frac{5}{2})}(z) &=&
\frac{\sqrt{3}}{2\sqrt{(N-1)(2N-1)(3N-2)}}
\Biggr[
2i\, f^{abc} \psi^a K^b L^c 
-\frac{2(N-1) i}{(N-2)}\, f^{abc} \psi^a
J^b L^c 
\nonu \\
& + & \frac{4(N-1) }{(N-2)}\, f^{abc} \psi^a \psi^b \psi^c J
-\frac{4(N-1) }{(N-2)}\,
\psi^a J^a J
-i\,\psi^a\,( d^{ajkb} \psi^j \chi^k J^b)
\nonu \\
& + & i\,\chi^i\,( d^{ibjc} \psi^b \psi^j K^c)\Biggr](z),
\label{GW5h}
\eea
where the numerator and denominator of relative coefficients behave linearly in $N$. Therefore, the lowest ${\cal N}=2$ higher spin multiplet in terms of two fermions in the coset model is given by (\ref{GW3h}) and (\ref{GW5h}) for general $N$. 
Section $4.2$
describes the OPEs between them. To obtain the next ${\cal N}=2$ higher spin multiplet for generic $N$, we calculate the OPEs between (\ref{GW3h}) and (\ref{GW5h}) (not presented here for space considerations). 


%
\subsection{ Operator product expansion between the first ${\cal N}=2$ higher spin   multiplet
}

The (super) OPE ${\bf W}^{(\frac{3}{2})}_{0}(Z_1) \, {\bf   W}^{(\frac{3}{2})}_{0}(Z_2)$ is completely determined from the Jocobi identity between the two (higher spin) currents  $({\bf T},{\bf W}^{(\frac{3}{2})}_{0}, {\bf W}^{(\frac{3}{2})}_{0})$, i.e., all the structure constants in the right hand side of the OPE are fixed in terms of the central charge or $N$, for general $N$~\cite{KT},
\bea
{\bf W}^{(\frac{3}{2})}_{0}(Z_1) \, {\bf
W}^{(\frac{3}{2})}_{0}(Z_2) & = & \frac{1}{z_{12}^3} \, \frac{2}{3}c
+\frac{\theta_{12} \bar{\theta}_{12}}{z_{12}^3} \,\, 3{\bf T}(Z_2)
\nonu \\
& + & \frac{1}{z_{12}^2} \Bigg[ -3\theta_{12} D {\bf T} +
3\bar{\theta}_{12}\overline{D} {\bf T} +3\theta_{12}
\bar{\theta}_{12} \pa {\bf T}
 \Bigg](Z_2)
\nonu \\
& + & \frac{1}{z_{12}} \frac{1}{(1-c)} \Bigg[ c [D, \overline{D}]{\bf
T} +3 {\bf T}{\bf T}
 \Bigg](Z_2)
\nonu \\
&+& \frac{\theta_{12}}{z_{12}} \frac{1}{(1-c)}
 \Bigg[
(2c-3) \pa D {\bf T} +3 {\bf T} D {\bf T}
\Bigg](Z_2)
\nonu \\
&+& \frac{\bar{\theta}_{12}}{z_{12}} \frac{1}{(1-c)}
\Bigg[ -(2c-3)\pa
\overline{D} {\bf T} +3 {\bf T}\overline{D} {\bf T}
\Bigg](Z_2)
+ \frac{\theta_{12} \bar{\theta}_{12}}{z_{12}} \Bigg[ C_{(\frac{3}{2})(\frac{3}{2})}^{(3)}{\bf
    W}^{(3)}_{0} \nonu \\
  & + & \frac{1}{(-1+c) (6+c) (-3+2 c)}
 \Bigg(-\frac{9}{4} c (-9+4 c) \pa [D, \overline{D}]  {\bf T}
 \nonu \\
&-& \frac{9}{2} (9-3 c+2 c^2) {\bf T} [D,
\overline{D} ] {\bf T}
- \frac{27}{2} (1+2 c) {\bf T} {\bf T} {\bf
  T} \nonu \\
& - & 9 c (-9+4 c) \overline{D}{\bf T}
D{\bf T}
+ \frac{3}{2}(18-18 c+3 c^2+2 c^3) \pa^2 {\bf
  T}\Bigg) \Bigg](Z_2) \nonu \\
& + & \cdots. \label{GsuperopeW3hW3h}
\eea 
Here the structure constant
\bea
(C_{(\frac{3}{2})(\frac{3}{2})}^{(3)})^2=
\frac{3(3 + 2 c) (-9 + 4 c) (-3 + 5 c)}{2(-1 + c)
    (6 + c) (-3 + 2 c)},
\label{cstructure}
\eea
which is fixed by requiring  that the sixth-order pole of OPE $W_{0}^{(3)}(z) \, W_{0}^{(3)}(w)$ should behave as $\frac{1}{(z-w)^6}\,\frac{c}{3}$, and we can express (\ref{cstructure}) in terms of $N$ using (\ref{centralcharge}). Under the infinity limit of $c$, (\ref{cstructure}) is finite and equals $30$. We can rewrite (\ref{GsuperopeW3hW3h}) in terms of various quasi primary fields as discussed in Section $3$ for the component approach.



From (\ref{GsuperopeW3hW3h}), OPEs between component fields of  ${\bf W}_{0}^{(\frac{3}{2})}(Z)$\footnote{$ W_{0}^{(\frac{3}{2})} =   {\bf W}_{0}^{(\frac{3}{2})}|_{\theta=\bar{\theta}=0} $, $ W_{+}^{(2)} =   D {\bf W}_{0}^{(\frac{3}{2})}|_{\theta=\bar{\theta}=0}$, $ W_{-}^{(2)} =   \overline{D} {\bf W}_{0}^{(\frac{3}{2})}|_{\theta=\bar{\theta}=0}$   and $ W_{0}^{(\frac{5}{2})} =   -\frac{1}{2}   [D, \overline{D}] {\bf W}_{0}^{(\frac{3}{2})}|_{\theta=\bar{\theta}=0}$. Similarly, $ J = {\bf T}|_{\theta=\bar{\theta}=0}$, $G^{+} = D {\bf T}|_{\theta=\bar{\theta}=0}$,$G^{-} = \overline{D} {\bf T}|_{\theta=\bar{\theta}=0}$ and $T = -\frac{1}{2} [D, \overline{D}] {\bf T}|_{\theta=\bar{\theta}=0}$ are also satisfied. Thus, we can obtain (\ref{opeW5hW3h}) by selecting $\theta_{12}$ and $\bar{\theta}_{12}$ independent terms, $\theta_{12}$ dependent terms, $\bar{\theta}_{12}$ dependent terms, and the remaining terms respectively, applying $D_1$ or $\overline{D}_1$ on both sides and putting $\theta_{12}=\bar{\theta}_{12}=0$. } can be expressed as 
\bea
W_{0}^{(\frac{3}{2})}(z) \, W_{0}^{(\frac{3}{2})}(w) & = &
\frac{1}{(z-w)^3} \, \frac{2c}{3} +
\frac{1}{(z-w)} \,\frac{1}{(1-c)} \Bigg[-2c\,T+3\,JJ \Bigg](w)
+\cdots,
\nonu \\
& \longrightarrow &
\frac{1}{(z-w)^3} \, \frac{2c}{3} +
\frac{1}{(z-w)} \, \Bigg[2\,T-\frac{3}{c}\,JJ \Bigg](w)
+\cdots,
\nonu \\
W_{\pm}^{(2)}(z) \,W_{0}^{(\frac{3}{2})}(w)  & = & 
\mp \frac{1}{(z-w)^2} \,3G^{\pm}(w) \nonu \\
& + &
\frac{1}{(z-w)} \, \,\frac{1}{(1-c)} \Bigg[3\,JG^{\pm} \pm
  (2 c-3) \pa G^{\pm} \Bigg](w)+\cdots,
\nonu 
\\
&\longrightarrow &
\mp \frac{1}{(z-w)^2} \,3G^{\pm}(w) + 
\frac{1}{(z-w)} \, \, \Bigg[ -\frac{3}{c}\,JG^{\pm} \mp
  2  \pa G^{\pm} \Bigg](w)+\cdots,
\nonu \\
W_{0}^{(\frac{5}{2})}(z) \, W_{0}^{(\frac{3}{2})}(w) & = &
\frac{1}{(z-w)^3}
 \, 3 J(w) + \frac{1}{(z-w)^2} \,  3 \pa J(w)
+
\frac{1}{(z-w)} \, \Bigg[
  C_{(\frac{3}{2})(\frac{3}{2})}^{(3)}W_{0}^{(3)}
  \nonu \\
  & + & \frac{1}{(-1+c) (6+c) (-3+2 c)}
  \Bigg(-
  9 c (-9+4 c) G^{-}G^{+}
\nonu \\
&+&9 (9-3c+2c^2) JT-
\frac{27}{2} (1+2 c)JJJ
\nonu \\
&+&\frac{9}{2} c (-9+4c) \pa T
+\frac{3}{2}(18-18 c+3 c^2+2 c^3)
\pa^2 J \Bigg) \Bigg](w)+ 
\cdots,
\nonu \\
& \longrightarrow & \frac{1}{(z-w)^3}
 \, 3 J(w) + \frac{1}{(z-w)^2} \,  3 \pa J(w)
\nonu \\
&+&
\frac{1}{(z-w)} \, \Bigg[
  C_{(\frac{3}{2})(\frac{3}{2})}^{(3)}W_{0}^{(3)}-
  \frac{18}{c}G^{-}G^{+}
+\frac{9}{c}JT-
\frac{27}{2c^2}JJJ
+\frac{3}{2} \pa^2 J \Bigg](w)\nonu \\
& + &
\cdots,
\label{opeW5hW3h}
\eea
where the large $c$ limit is taken by keeping $\frac{1}{c}$ for the quadratic fields and $\frac{1}{c^2}$ for cubic fields in the right hand sides of the OPEs in (\ref{opeW5hW3h})\footnote{\label{mode} In terms of mode expansions, 
  \bea
  \{W^{(\frac{3}{2})}_{r}, W^{(\frac{3}{2})}_s \}  & = &
  2 L_{r+s} + \frac{c}{3} (r^2-\frac{1}{4}) \delta_{r,-s},
  \nonu \\
  \big[ W^{(2) \pm}_{m}, W^{(\frac{3}{2})}_r\big] & = &
  \mp (m-2r) G^{\pm}_{m+r},
 \nonu \\
 \{W^{(\frac{5}{2}) }_{r}, W^{(\frac{3}{2})}_s \} & = &
 C^{(3)}_{(\frac{3}{2})
   (\frac{3}{2})} W^{(3)}_{r+s} + \frac{3}{8} (2s+1)(2s-1)J_{r+s},
\label{commanticomm}
\eea
up to modes from nonlinear terms, i.e., the infinity limit of $c$.  The central term vanishes for $r=\pm \frac{1}{2}$, and the second term of the last equation vanishes for $s =\pm \frac{1}{2}$.} (see~\cite{Ahn1206,Ahn1208} for the large $c$ limit). Thus, this classical algebra provides asymptotic symmetry algebra for the (matrix extension of) $AdS_3$ bulk theory~\cite{CHR}, and
Section $5$ describes the corresponding wedge algebra and Appendix $C$ provides the remaining OPEs explicitly.

\subsection{ Operator product expansions between first and second   ${\cal N}=2$ higher spin   multiplets
}

From the Jocobi identity between the three (higher spin) currents   $({\bf T},{\bf W}^{(\frac{3}{2})}_{0}, {\bf W}^{(3)}_{0})$,  and the Jocobi identity of the higher spin current   $({\bf W}^{(\frac{3}{2})}_{0},{\bf W}^{(\frac{3}{2})}_{0}, {\bf W}^{(\frac{3}{2})}_{0})$, the (super) OPE ${\bf W}^{(\frac{3}{2})}_{0}(Z_1) \, {\bf    W}^{(3)}_{0}(Z_2)$ for general $N$  can be expressed as 
 \bea
{\bf W}^{(\frac{3}{2})}_{0}(Z_1) \, {\bf
W}^{(3)}_{0}(Z_2) & = & \frac{\theta_{12} \bar{\theta}_{12}}{z_{12}^4}
\,\,\Bigg[\frac{3 (-9+4 c) (-9+9 c+10 c^2)}{4 (-1+c) (6+c) (-3+2
c)C^{(3)}_{(\frac{3}{2})(\frac{3}{2})}}\Bigg]{\bf W}^{(\frac{3}{2})}_{0}(Z_2)
\nonu \\
& + & \frac{1}{z_{12}^3} \,
\frac{(-9+4 c) (-9+9 c+10 c^2)}{2 (-1+c) (6+c) (-3+2 c)C^{(3)}_{(\frac{3}{2})(\frac{3}{2})}}
 \Bigg[\theta_{12}(-1) D {\bf W}^{(\frac{3}{2})}_{0}
   + \bar{\theta}_{12} \overline{D} {\bf W}^{(\frac{3}{2})}_{0} \Bigg](Z_2)
 \nonu \\
 & + &
\frac{\theta_{12} \bar{\theta}_{12}}{z_{12}^3}  \frac{1}{3}
 \pa (\mbox{pole-4})(Z_2)
 \nonu \\
 & + &  \frac{1}{z_{12}^2} \, 
  \Bigg[\frac{c (9-9 c-10
      c^2)}{2 (-1+c) (6+c) (-3+2 c) C^{(3)}_{(\frac{3}{2})(\frac{3}{2})}}
    [D, \overline{D}] {\bf
      W}^{(\frac{3}{2})}_{0} \nonu \\
    &- & \frac{9 (-9+9 c+10 c^2)}{2 (-1+c) (6+c) (-3+2 c)
      C^{(3)}_{(\frac{3}{2})(\frac{3}{2})}} {\bf T} {\bf W}^{(\frac{3}{2})}_{0}
    +  \theta_{12} \Bigg(
    \frac{1}{4}\pa (\mbox{pole-3})_{\bar{\theta}=0}
    +{\bf Q}_{1} \Bigg)
\nonu \\
& + &  \bar{\theta}_{12}
\Bigg( \frac{1}{4} \pa (\mbox{pole-3})_{{\theta}=0}
+ {\bf Q}_{2} \Bigg)
\nonu \\
&+&
\theta_{12}\bar{\theta}_{12} \Bigg( -\frac{7}{2}
 C^{(\frac{7}{2})}_{(\frac{3}{2})(3)}
{\bf W}^{(\frac{7}{2})}_{0}
+\frac{1}{12} \pa^2 (\mbox{pole-4})
+ {\bf Q}_{3} \Bigg)
 \Bigg](Z_2)
 \label{secondOPE} \\
& + & \frac{1}{z_{12}} \,
 \Bigg[ \frac{1}{5} \pa (\mbox{pole-2})_{\theta=\bar{\theta}=0}
   +{\bf Q}_{4}
\nonu \\
&+&  \theta_{12} \Bigg(  C^{(\frac{7}{2})}_{(\frac{3}{2})(3)}
D {\bf W}^{(\frac{7}{2})}_{0}+ \frac{1}{20} \pa^2
(\mbox{pole-3})_{\bar{\theta}=0}
+\frac{1}{3} \pa {\bf Q}_{1}+ {\bf Q}_{5} \Bigg)
\nonu \\
&+&   \bar{\theta}_{12} \Bigg(- C^{(\frac{7}{2})}_{(\frac{3}{2})(3)}
\bar{ D }{\bf W}^{(\frac{7}{2})}_{0}+ \frac{1}{20}
\pa^2 (\mbox{pole-3})_{\theta=0}
+\frac{1}{3}   \pa {\bf Q}_{2} + {\bf Q}_{6}\Bigg)
\nonu \\
&+&
\theta_{12} \bar{\theta}_{12} \Bigg(
-\frac{3}{2}  C^{(\frac{7}{2})}_{(\frac{3}{2})(3)}
\pa {\bf W}^{(\frac{7}{2})}_{0}
+\frac{1}{60} \pa^3 (\mbox{pole-4})
+\frac{3}{7} \pa {\bf Q}_{3}
+ {\bf Q}_{7} \Bigg)
\Bigg](Z_2) +\cdots,
 \nonu
\eea
where the structure constant is determined using simplified notations (e.g. in the expression of   $(\mbox{pole-3})_{\bar{\theta}=0}$ of (\ref{secondOPE}),   we take the third order pole of   (\ref{secondOPE}) by taking $\bar{\theta}=0$, leaving the first term of the third order pole) as
\bea
(C_{(\frac{3}{2})(3)}^{(\frac{7}{2})})^2=
\frac{(108 - 144 c + 15 c^2 + 7 c^3)}{
  (39 - 53 c + 13 c^2 + c^3)}
\label{cstruc}
\eea
by requiring that  the seventh-order pole for OPE $W_{0}^{({\frac{7}{2}})}(z) \, W_{0}^{({\frac{7}{2}})}(w)$ should have $\frac{1}{(z-w)^7}\,\frac{2c}{7}$.

Under the infinity limit of $c$, the structure constant (\ref{cstruc}) is finite and equals $7$.
The appearance of ${\bf   W}^{(\frac{7}{2})}_{0}(Z_2)$ in (\ref{secondOPE}) was also observed in (\ref{W5hW3}) for the component approach. The seven quasi primary fields, ${\bf Q}_1(Z_2), \cdots, {\bf Q}_7(Z_2)$, are presented in Appendix $D$. The large $c$ limit can be taken in (\ref{secondOPE}) as in Section $4.2$, but are not included here for space considerations. The nonlinear term in the second order pole in (\ref{secondOPE}) vanishes in this limit and the quasi primary fields in Appendix $D$ also vanish. We can then obtain the (anti)commutators from (\ref{secondOPE}), similarly to (\ref{commanticomm}) and (\ref{wremain}) (not presented here), which  provide the correct relations between higher spin generators corresponding to the first two ${\cal N}=2$ higher spin multiplets.

The two important results in this section are summarized by (\ref{GsuperopeW3hW3h}) and (\ref{secondOPE}), and other OPEs are given in Appendix $E$.

  \section{ $AdS_3$ higher spin theory with matrix generalization}

  We construct Vasiliev's oscillator description     corresponding to the first two ${\cal N}=2$ higher spin multiplet   discussed in previous Sections $3$ and $4$.
  
  \subsection{Wedge algebra for ${\cal N}=2$ superconformal algebra}

Lie algebra $\mbox{shs}[\la=\frac{1}{2}]$ is generated by   $\hat{y}_{\alpha} (\alpha=1,2)$ with fundamental commutator   $[\hat{y}_{\alpha}, \hat{y}_{\beta}] = 2 i   \ep_{\alpha \beta}$~\cite{PV9806,PV9812,BVdplb,BVdnpb,AKP}, which has no oscillator $k$ dependence.   Chan-Paton factors are introduced   and generators $\mbox{shs}_{M=2}[\la=\frac{1}{2}]$   are given by the tensor product between generators of $\mbox{shs}[\la=\frac{1}{2}]$ higher spin algebra   and $GL(2)$ generators. The 't Hooft parameter $\la =   \frac{2N}{(2N+k-1)}$ in~\cite{CHR,AP1310}   becomes $\frac{N}{(2N-1)}$ at   the critical level and tends to $\la =\frac{1}{2}$ under the   infinity limit of $N$.

Spin-$\frac{3}{2}$ currents of ${\cal N}=2$ superconformal   algebra play the role of the four fermionic generators for ${\cal N}=2$ wedge algebra~\cite{CHR1406} (${\cal N}=2$ truncation of ${\cal N}=4$   theory~\cite{GG1305}),
  \bea
&&G^{+}_{\frac{1}{2}} = (-\frac{i}{4})^{\frac{1}{2}}
\hat{y}_1  \otimes 
  \left( {\begin{array}{cc}
   0 & 1 \\
   0 & 0 \\
  \end{array} } \right) \, ,\qquad 
G^{+}_{-\frac{1}{2}} = (-\frac{i}{4})^{\frac{1}{2}}
\hat{y}_2  \otimes 
  \left( {\begin{array}{cc}
   0 & 1 \\
   0 & 0 \\
  \end{array} } \right), 
\nonu \\
&&G^{-}_{\frac{1}{2}} = (-\frac{i}{4})^{\frac{1}{2}}
\hat{y}_1 \otimes 
  \left( {\begin{array}{cc}
   0 & 0 \\
   1 & 0 \\
  \end{array} } \right) \, ,\qquad 
G^{-}_{-\frac{1}{2}} = (-\frac{i}{4})^{\frac{1}{2}}
\hat{y}_2  \otimes 
  \left( {\begin{array}{cc}
   0 & 0 \\
   1 & 0 \\
  \end{array} } \right). 
\label{fourg}
  \eea
Calculating the anticommutators between these operators in (\ref{fourg}), 
\bea
\{G^{+}_{r}, G^{-}_s \} =  L_{r+s} + \frac{1}{2}(r-s) J_{r+s},
\label{gg}
\eea
where the spin-$2$ current of ${\cal N}=2$ superconformal algebra plays the role of three bosonic generators, i.e., the matrix generalization of the ${\cal N}=2$ wedge algebra~\cite{GG1305},
\bea
&&L_{1} = (-\frac{i}{4})
\hat{y}_1 \hat{y}_1 \otimes 
  \left( {\begin{array}{cc}
   1 & 0 \\
   0 & 1 \\
  \end{array} } \right) \, ,\qquad 
L_{-1} = (-\frac{i}{4})
\hat{y}_2 \hat{y}_2 \otimes 
  \left( {\begin{array}{cc}
   1 & 0 \\
   0 & 1 \\
  \end{array} } \right), 
\nonu \\
&&L_{0} = (-\frac{i}{4}) \frac{1}{2}
(\hat{y}_1 \hat{y}_2 + \hat{y}_2 \hat{y}_1) \otimes 
  \left( {\begin{array}{cc}
   1 & 0 \\
   0 & 1 \\
  \end{array} } \right), 
\label{threel}
  \eea
and spin-$1$ has matrix form
\bea
J_{0} = \frac{1}{2} \otimes 
  \left( {\begin{array}{cc}
   1 & 0 \\
   0 & -1 \\
  \end{array} } \right).
\label{j}
  \eea
  We can check\footnote{ \label{two}     The remaining nontrivial nonzero     commutators are given by $[L_m, L_n]=(m-n) L_{m+n}$, $[L_m,     G_r^{\pm}] = (\frac{m}{2}-r) G^{\pm}_{m+r}$.}  the following commutators
  \bea
\left[ J_{0}, G^\pm_{r} \right] =\pm G^\pm_{r}.
\label{jcharge}
\eea
Thus, the ${\cal N}=2$ wedge algebra~\cite{GG1305} (generated by four bosonic and fermionic generators) is described by (\ref{gg}), (\ref{jcharge}), and those in footnote \ref{two}, together with (\ref{fourg}), (\ref{threel}) and (\ref{j}). Finally, ${\cal N}=2$ wedge algebra is reproduced by restricting the mode indices in Appendix $A$ to wedge cases.

  \subsection{ $OSp(2|2)$ higher spin algebra}

  From the classical limit discussed in Section $4$, we would like   to construct the higher spin generators corresponding to   the first two ${\cal N}=2$ higher spin multiplets.
  
  \subsubsection{Twelve higher spin generators}

We can obtain the (lowest) two higher spin generators by adding the matrix degree of freedoms to the bulk theory~\cite{CHR} and requiring that $[J_0, V^{(\frac{3}{2})}_{r}]=0$ and $[L_m, V^{(\frac{3}{2})}_{r}]= [(\frac{3}{2}-1)m-r] V^{(\frac{3}{2})}_{m+r}$ given in Appendix $B$,
\bea
&&V^{(\frac{3}{2})}_{\frac{1}{2}} = \frac{1}{2}
(-\frac{i}{4})^{\frac{1}{2}}
\hat{y}_1  \otimes 
  \left( {\begin{array}{cc}
   1 & 0 \\
   0 & 1 \\
  \end{array} } \right) \, ,\qquad 
V^{(\frac{3}{2})}_{-\frac{1}{2}} = \frac{1}{2} (-\frac{i}{4})^{\frac{1}{2}}
\hat{y}_2  \otimes 
  \left( {\begin{array}{cc}
   1 & 0 \\
   0 & 1 \\
  \end{array} } \right) ,
\label{vspin3half}
  \eea
which have the anticommutator
  \bea
&& \{V^{(\frac{3}{2})}_{r}, V^{(\frac{3}{2})}_s \} =
\frac{1}{2} L_{r+s}
\label{ope11}
\eea
corresponding to the first one in footnote~\ref{mode} because half of $W^{(\frac{3}{2})}_{r}$ satisfies (\ref{ope11}) by restricting mode $r$ to $\pm \frac{1}{2}$.

We can obtain the three higher spin generators by calculating the anticommutators $\{ G_r^{+}, V^{(\frac{3}{2})}_{s}\}=  -V^{(2)+}_{r+s}$ described in Appendix $B$ using (\ref{fourg}) and (\ref{vspin3half}), 
\bea
&&V_{1}^{(2)+} = -(-\frac{i}{4})
\hat{y}_1 \hat{y}_1 \otimes 
  \left( {\begin{array}{cc}
  0 & 1 \\
   0 & 0 \\
  \end{array} } \right) \, ,\qquad 
V_{-1}^{(2)+} = -(-\frac{i}{4})
\hat{y}_2 \hat{y}_2 \otimes 
  \left( {\begin{array}{cc}
    0 & 1 \\
   0 & 0 \\
  \end{array} } \right), 
\nonu \\
&&V_{0}^{(2)+} = -(-\frac{i}{4}) \frac{1}{2}
(\hat{y}_1 \hat{y}_2 + \hat{y}_2 \hat{y}_1) \otimes 
  \left( {\begin{array}{cc}
    0 & 1 \\
   0 & 0 \\
  \end{array} } \right). 
\label{vspin2+}
  \eea
Similarly, the three higher spin generators are
\bea
&&V_{1}^{(2)-} = (\frac{-i}{4})
\hat{y}_1 \hat{y}_1 \otimes 
  \left( {\begin{array}{cc}
  0 & 0 \\
   1 & 0 \\
  \end{array} } \right) \, ,\qquad 
V_{-1}^{(2)-} = (-\frac{i}{4})
\hat{y}_2 \hat{y}_2 \otimes 
  \left( {\begin{array}{cc}
     0 & 0 \\
   1 & 0 \\
  \end{array} } \right), 
\nonu \\
&&V_{0}^{(2)-} = (-\frac{i}{4}) \frac{1}{2}
(\hat{y}_1 \hat{y}_2 + \hat{y}_2 \hat{y}_1) \otimes 
  \left( {\begin{array}{cc}
     0 & 0 \\
   1 & 0 \\
  \end{array} } \right), 
\label{vspin2-}
  \eea
by   calculating the anticommutators  $\{ G_r^{-}, V^{(\frac{3}{2})}_{s}\}=  V^{(2)-}_{r+s}$  (see Appendix $B$)  using (\ref{fourg}) and (\ref{vspin3half}).  Thus, we obtain commutators
 \bea
     [ V_m^{(2) \pm}, V_s^{(\frac{3}{2})} ] = \mp
     \frac{1}{4} (m-2s) G^{\pm}_{m+s},
 \label{ope231}
 \eea
using (\ref{vspin2+}), (\ref{vspin2-}), (\ref{vspin3half}), and (\ref{fourg})\footnote{We have nontrivial relationships $ \left[ J_{0}, V^{(2) \pm}_{m} \right] =\pm V^{(2) \pm}_{m}$ (described in Appendix $B$), similar to (\ref{jcharge}).}.  Multiplying $\frac{1}{4}$ of the second  relationship (half of $W_m^{(2) \pm}$ and half of $W_r^{(\frac{3}{2})}$) in footnote~\ref{mode}, we can obtain   (\ref{ope231}) by restricting mode indices to wedge cases.
 
 Finally, from $[G^{\pm}_{r}, V^{(2)\mp}_m ] =  V^{(\frac{5}{2})}_{r+m}\pm \frac{1}{2} (2r-m) V^{(\frac{3}{2})}_{r+m}$  (see Appendix $B$),  we have the four higher spin generators
\bea
V^{(\frac{5}{2})}_{\frac{3}{2}}  & = & (-\frac{i}{4})^{\frac{3}{2}}
\hat{y}_1 \hat{y}_1 \hat{y}_1 \otimes 
  \left( {\begin{array}{cc}
  1 & 0 \\
   0 & -1 \\
  \end{array} } \right), \qquad
V^{(\frac{5}{2})}_{-\frac{3}{2}} = (-\frac{i}{4})^{\frac{3}{2}}
\hat{y}_2 \hat{y}_2 \hat{y}_2 \otimes 
  \left( {\begin{array}{cc}
  1 & 0 \\
   0 & -1 \\
  \end{array} } \right), \nonu \\  
  V^{(\frac{5}{2})}_{\frac{1}{2}}  & = &
  (-\frac{i}{4})^{\frac{3}{2}}
  \frac{1}{3}(\hat{y}_1 \hat{y}_1 \hat{y}_2 +
    \hat{y}_1 \hat{y}_2 \hat{y}_1 +
    \hat{y}_2 \hat{y}_1 \hat{y}_1 ) 
  \otimes \left( {\begin{array}{cc}
    1 & 0 \\
   0 & -1 \\
  \end{array} } \right), 
\nonu \\
V^{(\frac{5}{2})}_{-\frac{1}{2}} & = & (-\frac{i}{4})^{\frac{3}{2}}
\frac{1}{3}
(\hat{y}_1 \hat{y}_2 \hat{y}_2 + \hat{y}_2 \hat{y}_1 \hat{y}_2
+ \hat{y}_2 \hat{y}_2 \hat{y}_1
) 
 \otimes  \left( {\begin{array}{cc}
    1 & 0 \\
   0 & -1 \\
  \end{array} } \right).  
\label{vspin5half}
\eea
The matrices appearing in the higher spin generators for spins $\frac{3}{2}$ and $\frac{5}{2}$ are diagonal in (\ref{vspin3half}) and (\ref{vspin5half}). Decomposition of these in the above commutator can be achieved by moving the oscillators appropriately. Thus, we have nontrivial   relationships $\{G^{\pm}_{r}, V^{(\frac{5}{2})}_s \} =   \frac{1}{2} (3r-s)V^{(2)\pm}_{r+s}$, which also appear in   Appendix $B$.   There are several ways to express the last two higher spin generators of (\ref{vspin5half}), for convenience we express them symmetrically in the indices.
  
  \subsubsection{Next twenty-four higher spin generators}
  
We can calculate the anticommutators for the second ${\cal N}=2$ higher spin multiplet, together with (\ref{vspin3half}) and (\ref{vspin5half}) by recalling that higher spin-$3$ current appears in the OPE between higher spin-$\frac{5}{2}$ and spin-$\frac{3}{2}$ currents from Section $4.2$,
\bea
\{V^{(\frac{5}{2}) }_{r}, V^{(\frac{3}{2})}_s \} = V^{(3)}_{r+s}.
\label{ope41}
\eea
Compared to the previous (\ref{ope11}) and (\ref{ope231}), this result produces new higher spin generators in the right-hand side of (\ref{ope41}), 
\bea
V^{(3)}_{2} & = & (-\frac{i}{4})^{2}
\hat{y}_1 \hat{y}_1 \hat{y}_1 \hat{y}_1 \otimes 
  \left( {\begin{array}{cc}
  1 & 0 \\
   0 & -1 \\
  \end{array} } \right),
  \nonu \\
V^{(3)}_{1} & = & (-\frac{i}{4})^{2} \frac{1}{4}
(\hat{y}_1 \hat{y}_1 \hat{y}_1 \hat{y}_2 +
\hat{y}_1 \hat{y}_1 \hat{y}_2 \hat{y}_1 +
\hat{y}_1 \hat{y}_2 \hat{y}_1 \hat{y}_1 +
\hat{y}_2 \hat{y}_1 \hat{y}_1 \hat{y}_1 
) 
  \otimes \left( {\begin{array}{cc}
    1 & 0 \\
   0 & -1 \\
  \end{array} } \right),  
  \nonu \\
  V^{(3)}_{0} & = & (-\frac{i}{4})^{2} \frac{1}{6}
(\hat{y}_1 \hat{y}_1 \hat{y}_2 \hat{y}_2 +
  \hat{y}_1 \hat{y}_2 \hat{y}_1 \hat{y}_2+
\hat{y}_1 \hat{y}_2 \hat{y}_2 \hat{y}_1 +
\hat{y}_2 \hat{y}_1 \hat{y}_1 \hat{y}_2+
\hat{y}_2 \hat{y}_1 \hat{y}_2 \hat{y}_1 +
\hat{y}_2 \hat{y}_2 \hat{y}_1 \hat{y}_1
  ) 
  \otimes \left( {\begin{array}{cc}
    1 & 0 \\
   0 & -1 \\
  \end{array} } \right),  
  \nonu \\
V^{(3)}_{-1} & = & (-\frac{i}{4})^{2} \frac{1}{4}
(\hat{y}_2 \hat{y}_2 \hat{y}_2 \hat{y}_1 +
\hat{y}_2 \hat{y}_2 \hat{y}_1 \hat{y}_2+
\hat{y}_2 \hat{y}_1 \hat{y}_2 \hat{y}_2 +
\hat{y}_1 \hat{y}_2 \hat{y}_2 \hat{y}_2
) 
  \otimes \left( {\begin{array}{cc}
    1 & 0 \\
   0 & -1 \\
  \end{array} } \right),  
  \nonu  \\
V^{(3)}_{-2} & = & (-\frac{i}{4})^{2}
\hat{y}_2 \hat{y}_2 \hat{y}_2 \hat{y}_2 \otimes 
  \left( {\begin{array}{cc}
  1 & 0 \\
   0 & -1 \\
  \end{array} } \right).
  \label{nextlowest}
  \eea
  Following the approach in Section $5.2.1$,
  we present the generators symmetrically in the indices.   The anticommutators (\ref{ope41}) can be seen in the third relation of footnote~\ref{mode} with the wedge condition for the mode indices. We present the remaining generators for the second ${\cal N}=2$ higher spin multiplet in Appendix $F$\footnote{\label{remain}   We obtain the remaining (anti)commutators   using (\ref{vspin3half}), (\ref{vspin2+}), (\ref{vspin2-})   and (\ref{vspin5half}),
\bea
[V^{(2) -}_{m}, V^{(2) +}_{n} ] &= & V^{(3)}_{m+n}-\frac{1}{2}(m-n) L_{m+n}
+\frac{1}{8}(m^2-m n+n^2-1) J_{m+n},
\nonu \\
\big[V^{(\frac{5}{2})}_{r}, V^{(2) \pm}_{m} \big] &= & -
V^{(\frac{7}{2}) \pm}_{r+m}
+
\frac{1}{32}(9-4r^2+8r m-12m^2)G^{\pm}_{r+m},
\nonu \\
\{V^{(\frac{5}{2})}_{r}, V^{(\frac{5}{2})}_{s} \} &= & V^{(4)}_{r+s}
+
\frac{1}{16}(-9+6r^2-8rs+6s^2)L_{r+s}.
\label{remainingones}
\eea
Therefore, the $OSp(2|2)$ higher spin algebra (extension of ${\cal N}=2$ wedge algebra) contains (\ref{ope11}), (\ref{ope231}), (\ref{ope41}), and (\ref{remainingones}). We can then calculate the (anti)commutators between the higher spin generators described in Sections $5.2.1$ and $5.2.2$, and we expect they will appear in the corresponding (anti)commutators in Section $4.2$ with an appropriate limit. }. We can continue to determine the next generators corresponding to the third ${\cal N}=2$ higher spin multiplet using (\ref{secondOPE}), which will take the form of the tensor product of
the generators, i.e., symmetrized products of $\hat{y}_{\al}$'s  and $GL(2)$ generators. Although we can calculate their (anti)commutators, this is not presented here for space considerations.
  
  \section{Conclusions }

This paper analyzed the ${\cal N}=1$ holographic minimal model   at the critical level, determined the lowest ${\cal N}=2$   higher spin currents, and obtained their OPEs and OPEs   between these higher spin currents and second ${\cal N}=2$   higher spin currents. We also investigated Vasiliev's oscillator   construction with matrix degrees of freedom, which generalizes $OSp(2|2)$ superconformal   algebra.  

Several relevant problems remain open, which will be investigated in the near future.
\begin{itemize} 
\item The next ${\cal N}=2$ higher spin multiplet in terms of   two adjoint fermions.

The expression for higher spin currents for general $N$ was obtained for the lowest cases. It remains an open problem to   determine the next ${\cal N}=2$ higher spin currents for generic   $N$. Once the lowest component field of this   multiplet is obtained,   the remaining components can be determined using spin-$\frac{3}{2}$ currents in (\ref{deetee}) from Appendix $B$.
  
\item More general coset model.

Creutzig et al.~\cite{CHR1306} generalized coset (\ref{coset}), and it would be interesting to observe the presence of   the higher spin currents (see also~\cite{EGR,CH1812,KS1812} for recent relevant works   in different context).
  
\item ${\cal N}=2$ superspace description for the two   adjoint fermions.

We determined the lowest ${\cal N}=2$ higher spin currents   for general $N$ in Section~$4.1$, but it remains an open problem whether we can express the two adjoint fermions in ${\cal N}=2$   superspace explicitly, which would simplify some complicated   calculations.

\item General structure of ${\cal N}=2$ higher spin   algebra.

  We determined some structure constants for ${\cal N}=2$   higher spin algebra corresponding to the coset model (\ref{coset}) in Section $5.2$, but it remains an open problem to derive the complete set of structure constants~\cite{BVdplb,BVdnpb}.   Any (anti)commutators between the higher spin generators   consist of a sum of lower higher spin generators   as well as generators from ${\cal N}=2$ wedge algebra. The right-hand sides will be considerably simplified by considering the (anti)commutator
  with the $U(1)$ charges. However, the main task is how to express the structure constants   in terms of arbitrary modes and spins.
  
\item Vasiliev's oscillator formalism with matrix   generalization.

  It would be interesting to construct   the Vasiliev's oscillator formalism in different coset models~\cite{Ahn1211,Ahn1305,Ahn1604,AK1607,Ahn1701}.   We can analyze the present method in the general coset model, e.g. in the context of large ${\cal N}=4$ holography~\cite{GG1305}.   It may be difficult to construct OPEs   as spins increase in the coset model, but the   oscillator construction will assist with this and although the procedure to obtain   the (anti)commutators in terms of oscillators is rather tedious,    we can confirm the presence of higher spin generators   by counting the number of oscillators.
  
\item Asymptotic symmetry algebra at the quantum level.
  
We found only two (super) OPEs between   the first and second ${\cal N}=2$ higher spin multiplets in Section $4$, with some structure constants   appearing in the right hand sides of the remaining OPEs remaining undetermined (see Appendix $E$).   To find these, we need to obtain OPEs between   the higher spin currents in terms of two adjoint fermions. However, even the $N=4$ case leads to very complicated singular terms.   It would be interesting to obtain the complete OPEs from Appendix $E$   by determining all the unknown structure constants. 

From the behavior of higher spin algebra described in Section $5$,  anticommutators between generator   $V_0^{(\frac{5}{2})}$ of higher   spin-$\frac{5}{2}$ and generators for half integer   higher spin provide lowest component generators   corresponding to   the second, fourth, and sixth of (\ref{supermultiplets}), etc. (i.e., $V_0^{(3)}, V_0^{(5)},   V_0^{(7)}, \cdots$). On the other hand,    commutators between generator   $V_0^{(\frac{5}{2})}$ of higher   spin-$\frac{5}{2}$ and generators of integer   higher spin provide lowest component generators corresponding to    the third, fifth, seventh, etc. (i.e., $V_0^{(\frac{7}{2})},   V_0^{(\frac{11}{2})},   V_0^{(\frac{15}{2})}, \cdots$). The remaining undetermined structure   constants in Appendix $E$ will be fixed by applying Jacobi identities associated with these OPEs and assuming the OPEs between the first three   ${\cal N}=2$ higher spin multiplets in (\ref{supermultiplets})   and the fourth and the fifth multiplets (expressing the possible terms in the right hand sides).
\end{itemize}
\vspace{.7cm}

\centerline{\bf Acknowledgments}

CA would like to thank Y. Hikida for the many helpful discussions.
This research was supported by the Basic Science Research Program through the National Research Foundation of Korea   funded by the Ministry of Education   (grant. 2017R1D1A1A09079512). CA acknowledges the warm hospitality from  the School of  Liberal Arts(and Institute of Convergence Fundamental Studies, Seoul National University of Science and Technology.

\newpage

\appendix

\renewcommand{\theequation}{\Alph{section}\mbox{.}\arabic{equation}}

\section{ ${\cal N}=2$   superconformal algebra}

Nontrivial OPEs between the four ${\cal N}=2$ superconformal algebra currents~\cite{BFK} can be summarized as
\bea
J(z) \, J(w) & = & \frac{1}{(z-w)^2} \, \frac{c}{3} + \cdots,
\nonu \\
J(z) \, G^{+}(w) & = & \frac{1}{(z-w)} \, G^{+}(w) + \cdots,
\nonu \\
J(z) \, G^{-}(w) & = & -\frac{1}{(z-w)} \, G^{-}(w) + \cdots,
\nonu \\
J(z) \, T(w) & = & \frac{1}{(z-w)^2} \, J(w) + \cdots,
\nonu \\
G^{+}(z) \, G^{-}(w) & = & \frac{1}{(z-w)^3} \, \frac{c}{3} +
\frac{1}{(z-w)^2} \, J(w) +
\frac{1}{(z-w)} \, \left[ T + \frac{1}{2} \pa J \right](w) + \cdots,
\nonu \\
G^{+}(z) \, T(w) & = &
\frac{1}{(z-w)^2} \, \frac{3}{2} \, G^{+}(w) +
\frac{1}{(z-w)} \, \frac{1}{2} \, \pa G^{+}(w) +
\cdots,
\nonu \\
G^{-}(z) \, T(w) & = &
\frac{1}{(z-w)^2} \, \frac{3}{2} \, G^{-}(w) +
\frac{1}{(z-w)} \, \frac{1}{2} \, \pa G^{-}(w) +
\cdots,
\nonu \\
T(z) \, T(w) & = &
\frac{1}{(z-w)^4} \, \frac{c}{2} +\frac{1}{(z-w)^2} \, 2 T(w) +
\frac{1}{(z-w)} \, \pa T(w) +\cdots.
\label{n2scaexpression}
\eea
The ${\cal N}=2$ superspace description can be found in~\cite{Schoutens1988}, and for convenience we present the (anti)commutators, 
\bea
[ J_{m}, J_{n} ] & = & \frac{c}{3} m \delta_{m,-n},
\nonu  \\
\big[ J_{m}, G^\pm_{r} \big] & = & \pm G^\pm_{m+r}, 
\nonu  \\
\big[ J_{m}, L_{n} \big] & = &  m J_{m+n},
\nonu  \\
\{G^{+}_{r}, G^{-}_s \} & = &
L_{r+s} + \frac{1}{2}(r-s) J_{r+s} +
\frac{c}{6} (r^2-\frac{1}{4}) \delta_{r,-s},
\nonu  \\
\big[ G^{\pm}_{r}, L_{m} \big] & = &  (r-\frac{m}{2}) G^{\pm}_{r+m},
\nonu  \\
\big[ L_{m}, L_{n} \big] & = &
(m-n) L_{m+n} +  \frac{c}{12} (m^3-m) \delta_{m,-n}.
\label{n2scon}
\eea
The central terms in (\ref{n2scon}) vanish at $m=0$, $r =\pm \frac{1}{2}$, or $m=0 \pm 1$.

\section{Operator product expansions between the four currents  and higher spin currents}

Nontrivial OPEs between the four ${\cal N}=2$ superconformal algebra currents and four higher spin  currents $(W_q^{(h)},W_{q+1}^{(h+\frac{1}{2})},  W_{q-1}^{(h+\frac{1}{2})},  W_q^{(h+1)})$ with spins and $U(1)$ charges can be expressed as 
\bea
J(z) \,W_q^{(h)}(w) & = & \frac{1}{(z-w)} \, q \, W_q^{(h)}(w) + \cdots,
\nonu \\
J(z) \, W_{q+1}^{(h+\frac{1}{2})}(w) & = & \frac{1}{(z-w)} \, \left(q+1\right) \,
W_{q+1}^{(h+\frac{1}{2})}(w) + \cdots,
\nonu \\
J(z) \, W_{q-1}^{(h+\frac{1}{2})}(w) & = & \frac{1}{(z-w)} \, \left(q-1\right) \,
W_{q-1}^{(h+\frac{1}{2})}(w) + \cdots,
\nonu \\
J(z) \, W_q^{(h+1)}(w) & = & \frac{1}{(z-w)^2} \, h \, W_q^{(h)}(w)+
\frac{1}{(z-w)} \, q \,W_q^{(h+1)}(w) + \cdots,
\nonu \\
G^{+}(z) \, W_q^{(h)}(w) & = & -\frac{1}{(z-w)} \,  W_{q+1}^{(h+\frac{1}{2})}(w)
+ \cdots,
\nonu \\
G^{+}(z) \, W_{q-1}^{(h+\frac{1}{2})}(w) & = & \frac{1}{(z-w)^2} \,
\left(h + \frac{q}{2} \right) W_q^{(h)} + \frac{1}{(z-w)} \left[
W_{q}^{(h+1)} +\frac{1}{2} \pa W_q^{(h)} \right](w) + \cdots,
\nonu \\
G^{+}(z) \, W_q^{(h+1)}(w) & = & \frac{1}{(z-w)^2} \,
\left[h+\frac{1}{2} (q+1) \right] \,
W_{q+1}^{(h+\frac{1}{2})}(w)+
\frac{1}{(z-w)} \, \frac{1}{2}\, \pa  \,W_{q+1}^{(h+\frac{1}{2})}(w) + \cdots,
\nonu \\
G^{-}(z) \, W_q^{(h)}(w) & = & \frac{1}{(z-w)} \, W_{q-1}^{(h+\frac{1}{2})}(w)
+ \cdots,
\nonu \\
G^{-}(z) \, W_{q+1}^{(h+\frac{1}{2})}(w) & = & \frac{1}{(z-w)^2} \,
\left(-h + \frac{q}{2} \right) W_q^{(h)} + \frac{1}{(z-w)} \left[
W_{q}^{(h+1)} -\frac{1}{2} \pa W_q^{(h)} \right](w) + \cdots,
\nonu \\
G^{-}(z) \, W_q^{(h+1)}(w) & = & \frac{1}{(z-w)^2} \,
\left[h-\frac{1}{2} (q-1) \right] \,
W_{q-1}^{(h+\frac{1}{2})}(w)+
\frac{1}{(z-w)} \, \frac{1}{2}\, \pa  \, W_{q-1}^{(h+\frac{1}{2})}(w) + \cdots,
\nonu \\
T(z) \, W_q^{(h)}(w) & = & \frac{1}{(z-w)^2} \, h \, W_q^{(h)}(w) +
 \frac{1}{(z-w)} \, \pa \, W_q^{(h)}(w) + \cdots,
\nonu \\
T(z) \, W_{q+1}^{(h+\frac{1}{2})}(w) & = & \frac{1}{(z-w)^2} \, (h+\frac{1}{2}) \,
W_{q+1}^{(h+\frac{1}{2})}(w) +
 \frac{1}{(z-w)} \, \pa \, W_{q+1}^{(h+\frac{1}{2})}(w) + \cdots,
\nonu \\
T(z) \, W_{q-1}^{(h+\frac{1}{2})}(w) & = & \frac{1}{(z-w)^2} \, (h+\frac{1}{2}) \,
W_{q-1}^{(h+\frac{1}{2})}(w) +
 \frac{1}{(z-w)} \, \pa \, W_{q-1}^{(h+\frac{1}{2})}(w) + \cdots,
\nonu \\
T(z) \, W_q^{(h+1)}(w) & = & \frac{1}{(z-w)^3} \, \frac{q}{2} \, W_q^{(h)} +
 \frac{1}{(z-w)^2} \, (h+1) \, W_q^{(h+1)}(w) +
 \frac{1}{(z-w)} \, \pa \, W_q^{(h+1)}(w) \nonu \\
& + & \cdots.
\label{TWq}
\eea
The ${\cal N}=2$ superspace description can be found in~\cite{Ahn1604}, and for convenience, we present the (anti)commutators. The first four can be expressed as 
\bea
 [J_{m}, W^{(h,q)}_n] & = &  q\,W^{(h,q)}_{m+n},
\nonu \\
\big[J_{m}, W^{(h+\frac{1}{2},q \pm 1)}_n \big] & = &
(q \pm 1)\,W^{(h+\frac{1}{2},q \pm 1)}_{m+n},
\nonu \\
\big[J_{m}, W^{(h+1,q)}_n \big] & = &
q\,W^{(h+1,q)}_{m+n} +mh\,W^{(h,q)}_{m+n} .
\label{jcomm}
\eea
If $h$ is integer, the next six are equivalent to
\bea
[G^{\pm}_{r}, W^{(h,q)}_n ] & = &  \mp W^{(h+\frac{1}{2},q \pm 1)}_{r+n} ,
\nonu \\
\{G^{\pm}_{r}, W^{(h+\frac{1}{2},q \mp 1)}_s \} & = &
W^{(h+1,q )}_{r+s}+\Bigg[\frac{q}{2}(r+\frac{1}{2})\pm (h-
  \frac{1}{2})r \mp\frac{s}{2}\Bigg] W^{(h,q)}_{r+s} ,
\nonu \\
      \big[G^{\pm}_{r}, W^{(h+1,q)}_n \big] & = &
      \Bigg[(h r-\frac{n}{2}) \pm \frac{q}{2}(r+\frac{1}{2})\Bigg]
      W^{(h+\frac{1}{2},q \pm 1)}_{r+n} ;
\label{gcomm1}
\eea
and if $h$ is half-integer, then
\bea
 \{G^{\pm}_{r}, W^{(h,q)}_n \} & = &  \mp W^{(h+\frac{1}{2},q \pm 1)}_{r+n} ,
\nonu \\
\big[G^{\pm}_{r}, W^{(h+\frac{1}{2},q \mp 1)}_n \big] & = &
W^{(h+1,q )}_{r+n}+\Bigg[\frac{q}{2}(r+\frac{1}{2})\pm
  (h-\frac{1}{2})r \mp\frac{n}{2}\Bigg] W^{(h,q)}_{r+n} ,
\nonu \\
\{G^{\pm}_{r}, W^{(h+1,q)}_s \} & = &
\Bigg[(h r-\frac{s}{2}) \pm \frac{q}{2}(r+\frac{1}{2})\Bigg]
W^{(h+\frac{1}{2},q \pm 1)}_{r+s} .
\label{gcomm2}
\eea
We have the following commutators for the last four of (\ref{TWq})
\bea
[L_{m}, W^{(h,q)}_n] & = &  \Bigg[(h-1)m -n\Bigg]\,W^{(h,q)}_{m+n},
\nonu \\
\big[L_{m}, W^{(h+\frac{1}{2},q \pm 1)}_n \big] & = &
\Bigg[(h-\frac{1}{2})m -n\Bigg]\,W^{(h+\frac{1}{2},q \pm 1)}_{m+n},
\nonu \\
\big[L_{m}, W^{(h,q +1)}_n \big] & = &
(hm -n)\,W^{(h+1,q)}_{m+n} + \frac{q}{4} (m+1) m \,W^{(h,q)}_{m+n}.
\label{lcomm}
\eea
We can derive the $q=0$ case from (\ref{jcomm}), (\ref{gcomm1}), (\ref{gcomm2}), and (\ref{lcomm}).

\section{Remaining operator product expansions for Section $4.2$}

One way to determine  the remaining nontrivial OPEs for Section $4.2$  is to start with the OPE in (\ref{GsuperopeW3hW3h}), take super derivatives of both sides, and then apply $\theta_{12}=0$ and/or $\bar{\theta}_{12}=0$ constraints to extract OPE components. Following this process, the remaining nontrivial OPEs, including the two trivial cases, can be expressed as (at the large $c$ limit)
\bea
W_{+}^{(2)}(z) \, W_{-}^{(2)}(w) & = & -\frac{1}{(z-w)^4} c  -
\frac{1}{(z-w)^3}\,3J -  \frac{1}{(z-w)^2}\, \Bigg[
  -\frac{(3-4 c)}{(-1+c)}T\nonu \\
  & - &
  \frac{3}{2(-1+c)}JJ+\frac{3}{2} \pa J\Bigg](w)
-\frac{1}{(z-w)} \, \Bigg[C_{(\frac{3}{2})(\frac{3}{2})}^{(3)}W_{0}^{(3)}
  \nonu \\
  & + & \frac{1}{(-1+c) (6+c) (-3+2 c)}
  \Bigg(- 6(c-3) (5 c-3)  G^{-}G^{+}
\nonu \\
&+&
3 (9+8 c^2) JT
-\frac{27}{2} (1+2 c) JJJ
\nonu \\
&-& \frac{3}{2}(6+c) (-3+2 c) \pa J J
-\frac{1}{4} (108-207 c+60 c^2+8 c^3) \pa T
\nonu \\
&+&\frac{c}{2} (9-3 c+2 c^2)
\pa^2 J \Bigg) \Bigg](w)+\cdots,
\nonu
\\
& \longrightarrow&
-\frac{1}{(z-w)^4} c  -
\frac{1}{(z-w)^3}\,3J- \frac{1}{(z-w)^2}\, \Bigg[
  4 T-
  \frac{3}{2c} JJ+ \frac{3}{2}\pa J\Bigg](w)
\nonu \\
&-&\frac{1}{(z-w)} \, \Bigg[C_{(\frac{3}{2})(\frac{3}{2})}^{(3)}W_{0}^{(3)}
- \frac{15}{c} G^{-}G^{+}
+
\frac{12}{c} JT
-\frac{27}{2c^2}JJJ
\nonu \\
&
-& \frac{3}{2c} \pa J J
-\pa T
+\frac{1}{2}
\pa^2 J \Bigg](w)+\cdots,
\nonu \\
W_{0}^{(\frac{5}{2})}(z) \, W_{+}^{(2)}(w) & = & 
-\frac{1}{(z-w)^3} \, 6G^{+}(w) -
\frac{1}{(z-w)^2} \frac{1}{2(-1+c)} \, \Bigg[3\,JG^{+}-(9-8c) \pa G^{+} \Bigg](w)
\nonu \\
&-&
\frac{1}{(z-w)} \, \Bigg[C_{(\frac{3}{2})(\frac{3}{2})}^{(3)}
  W_{+}^{(\frac{7}{2})}
  + \frac{1}{(-1+c) (6+c) (-3+2 c)} \Bigg(
  27 (3-4 c+2 c^2) T G^{+}
\nonu \\
&-&\frac{81}{2} (1+2 c) JJG^{+}
-\frac{9}{2} c (-21+2 c) \pa G^{+} J
\nonu \\
&-&\frac{9}{2} (9-4 c) c \pa J  G^{+}
+\frac{3}{4} c (4 c^2-18 c+45)  \pa^2 G^{+} \Bigg)
\Bigg](w)+\cdots,
\nonu \\
&\longrightarrow&
-\frac{1}{(z-w)^3} \, 6G^{+}(w) -
\frac{1}{(z-w)^2}  \, \Bigg[\frac{3}{2c}\,JG^{+}+4 \pa G^{+} \Bigg](w)
\nonu \\
&-&
\frac{1}{(z-w)} \, \Bigg[C_{(\frac{3}{2})(\frac{3}{2})}^{(3)}
  W_{+}^{(\frac{7}{2})}
+\frac{27}{c} T G^{+}
- \frac{81}{2c^2} JJG^{+}
- \frac{9}{2c} \pa G^{+} J
\nonu \\
& + & \frac{9}{c} \pa J  G^{+}
+ \frac{3}{2} \pa^2 G^{+}\Bigg](w)+\cdots,
\nonu \\
W_{0}^{(\frac{5}{2})}(z) \, W_{-}^{(2)}(w) & = & 
-\frac{1}{(z-w)^3} \, 6G^{-}(w) +
\frac{1}{(z-w)^2} \frac{1}{2(-1+c)} \, \Bigg[3\,JG^{-}+(9-8c) \pa G^{-} \Bigg](w)
\nonu \\
&-&
\frac{1}{(z-w)} \, \Bigg[C_{(\frac{3}{2})(\frac{3}{2})}^{(3)}
  W_{-}^{(\frac{7}{2})}
+ \frac{1}{(-1+c) (6+c) (-3+2 c)}\Bigg( 27 (3-4 c+2 c^2) G^{-}T
\nonu \\
&-&\frac{81}{2} (1+2 c) JJG^{-}
+\frac{9}{2} c (-21+2 c)  \pa G^{-} J
\nonu \\
&+&\frac{9}{2} (9-4 c) c \pa J  G^{-}
+\frac{3}{4} (2 c+3) (2 c^2-3 c+9)  \pa^2 G^{-}
\Bigg)
\Bigg](w)+\cdots,
\nonu
\\
& \longrightarrow &
-\frac{1}{(z-w)^3} \, 6G^{-}(w) +
\frac{1}{(z-w)^2}  \, \Bigg[\frac{3}{2c}\,JG^{-}-4 \pa G^{-} \Bigg](w)
\nonu \\
&-&
\frac{1}{(z-w)} \, \Bigg[C_{(\frac{3}{2})(\frac{3}{2})}^{(3)}
  W_{-}^{(\frac{7}{2})}
+ \frac{27}{c} G^{-}T
- \frac{81}{2c^2}JJG^{-}
+ \frac{9}{2c} \pa G^{-} J
\nonu \\
& - & \frac{9}{c} \pa J  G^{-}
+ \frac{3}{2} \pa^2 G^{-}\Bigg](w)+\cdots,
\nonu \\
W_{0}^{(\frac{5}{2})}(z) \, W_{0}^{(\frac{5}{2})}(w)  & = &
\frac{1}{(z-w)^5} \,2c +
\frac{1}{(z-w)^3}\frac{1}{(-1+c)}\Bigg[-(9-10c)T -\frac{3}{2}
  JJ\Bigg](w)
\nonu \\
&+& \frac{1}{(z-w)^2}\, \frac{1}{(-1+c)}\Bigg[-
  \frac{1}{2}(9-10c)\pa T-\frac{3}{2}  \pa J J\Bigg](w)
\nonu \\
&+&\frac{1}{(z-w)} \, \Bigg[C_{(\frac{3}{2})(\frac{3}{2})}^{(3)}W_{0}^{(4)}
+ \frac{1}{(-1+c) (6+c) (-3+2 c)}\Bigg(27 (-5+c) c \pa G^{-} G^{+}
\nonu \\
&-&81 (1+2 c)J G^{-} G^{+}
-\frac{81}{2} (1+2 c) JJT
+\frac{81}{2} (1+2 c) \pa T J
\nonu \\
& + & 27 (3-4 c+2 c^2) TT
+27 (-5+c) c \pa G^{+} G^{-}
+\frac{9}{4} (9-4 c) c \pa J \pa J
\nonu \\
&+&\frac{9}{4} (-21+2 c) c \pa^2 J  J
+
\frac{3}{4} c (4 c^2-18 c+45) \pa^2 T
-\frac{27}{4} (1+2c) \pa^3 J \Bigg) \Bigg](w)
\nonu \\
& + & \cdots,
\nonu \\
& \longrightarrow&
\frac{1}{(z-w)^5} \,2c +
\frac{1}{(z-w)^3} \Bigg[ 10 T -\frac{3}{2c}
  JJ\Bigg](w)
\nonu \\
& + & \frac{1}{(z-w)^2}\, \Bigg[
  5 \pa T-\frac{3}{2c}  \pa J J\Bigg](w)
\label{opeW5hW5h}
\\
&+&\frac{1}{(z-w)} \, \Bigg[C_{(\frac{3}{2})(\frac{3}{2})}^{(3)}W_{0}^{(4)}
+ \frac{27}{2c} \pa G^{-} G^{+}
- \frac{81}{c^2} J G^{-} G^{+}
-\frac{81}{2c^2} JJT
\nonu \\
&+& \frac{81}{2c^2} \pa T J
+\frac{27}{c} TT
+ \frac{27}{2c} \pa G^{+} G^{-}
-\frac{9}{2c} \pa J \pa J
+ \frac{9}{4c} \pa^2 J  J
+
 \frac{3}{2} \pa^2 T \Bigg](w)+\cdots.
\nonu
\eea

To compare with the (anti)commutators from the oscillator description, we also need the commutators and anticommutators corresponding to (\ref{opeW5hW5h}),
\bea
[W^{(2) -}_{m}, W^{(2) +}_{n}] & = &
C^{(3)}_{(\frac{3}{2})(\frac{3}{2})}
W^{(3)}_{m+n}- 2(m-n) L_{m+n}
+\frac{1}{2}(m^2-mn+n^2-1) J_{m+n}
\nonu \\
&-&
 \frac{c}{6}(m+1)m(m-1)\delta_{m,-n},
\nonu \\
\big[W^{(\frac{5}{2})}_{r}, W^{(2) \pm}_{n}\big] &= &
-C^{(3)}_{(\frac{3}{2})(\frac{3}{2})} W^{(\frac{7}{2}) \pm}_{r+n}
+
\frac{1}{8}(9-4r^2+8rn-12n^2)G^{\pm}_{r+n},
\nonu \\
\{W^{(\frac{5}{2})}_{r},W^{(\frac{5}{2})}_{s} \} &= & 
C^{(3)}_{(\frac{3}{2})(\frac{3}{2})}
W^{(4)}_{r+s}
+
\frac{1}{4}(-9+6r^2-8rs+6s^2)L_{r+s}
\nonu \\
&+&
\frac{c}{12}(r-\frac{3}{2})(r-\frac{1}{2})
(r+\frac{1}{2})(r+\frac{3}{2})\delta_{r,-s},
\label{wremain}
\eea
where modes coming from nonlinear terms were ignored, i.e., the infinity limit of $c$.  The central terms vanish for $m=0, \pm 1$ or $r=\pm \frac{1}{2}, \pm \frac{3}{2}$. As above,  (\ref{wremain}) contain the cases in footnote~\ref{remain} by restricting mode indices to wedge cases. Thus, we have the complete (anti)commutators for (\ref{commanticomm}) and (\ref{wremain}) for the lowest ${\cal N}=2$ higher spin multiplet.

\section{ Quasi primary operators   from Section $4.3$  }


The various quasi primary fields from Section $4.3$ can be expressed as 
\bea
    {\bf Q}_{1} &=& \frac{1}{(-1+c) (6+c) (-3+2 c) C^{(3)}_{(\frac{3}{2})(\frac{3}{2})}} \Bigg[
      \frac{27}{8} (15-31 c+10 c^2) \pa D {\bf W}^{(\frac{3}{2})}_{0}
      \nonu \\
      & - & \frac{9}{2} (-3+5 c)(-3+2c)
       {\bf T}D {\bf W}^{(\frac{3}{2})}_{0}
+ 9 (9-18 c+5 c^2)
D{\bf T} {\bf W}^{(\frac{3}{2})}_{0} \Bigg],
\nonu \\
      {\bf Q}_{2} &=& \frac{1}{ (-1+c) (6+c) (-3+2 c) C^{(3)}_{(\frac{3}{2})(\frac{3}{2})}} \Bigg[-\frac{27}{8} (15-31 c+10 c^2)
          \pa \overline{ D } {\bf W}^{(\frac{3}{2})}_{0}
          \nonu \\
          & - & \frac{9}{2} (-3+5 c)(-3+2c)
            {\bf T}\overline{ D } {\bf W}^{(\frac{3}{2})}_{0}
+9 (9-18 c+5 c^2)
\overline{ D }{\bf T} {\bf W}^{(\frac{3}{2})}_{0} \Bigg],
\nonu \\
      {\bf Q}_{3} &=& \frac{1}{(-1+c) (6+c) (-3+2 c) (-39+14 c+c^2)
        C^{(3)}_{(\frac{3}{2})(\frac{3}{2})}} \nonu \\
      & \times & \Bigg[-\frac{
          9}{4} (27-387 c+585 c^2-31 c^3+10 c^4)
  {\bf T} [D,\overline{D}] {\bf W}^{(\frac{3}{2})}_{0}
  \nonu \\
  &-& \frac{9}{4} (-297+117 c+510 c^2+200 c^3)
  {\bf T} {\bf T}{\bf W}^{(\frac{3}{2})}_{0}
- \frac{9}{2} c (54-15 c-143 c^2+30 c^3)
           \overline{D} {\bf T} D{\bf W}^{(\frac{3}{2})}_{0}
\nonu \\
&+& \frac{1}{4}(5103-10773 c+3861 c^2+63 c^3-330 c^4)
          [D,\overline{D}]{\bf T} {\bf
W}^{(\frac{3}{2})}_{0}
\nonu \\
&+& \frac{9}{2} c (54-15 c-143 c^2+30 c^3)
       D {\bf T} \overline{D}{\bf W}^{(\frac{3}{2})}_{0}
 \nonu \\
 &-& \frac{9}{16} (-1701+3699 c-1317 c^2-307 c^3+170 c^4)
           \pa^2  {\bf W}^{(\frac{3}{2})}_{0} \Bigg],
\nonu \\
      {\bf Q}_{4} &=& \frac{1}{(-1+c) (6+c) (-3+2 c) C^{(3)}_{(\frac{3}{2})(\frac{3}{2})}} \Bigg[
        \frac{18}{5} c (-9+4 c)
           \pa  [D, \overline{D}] {\bf W}^{(\frac{3}{2})}_{0}
+\frac{27}{5} c (-9+4 c)   {\bf T}\pa  {\bf W}^{(\frac{3}{2})}_{0}
\nonu \\
&+&
9 c (-9+4 c)
\overline{D}{\bf T} D {\bf W}^{(\frac{3}{2})}_{0} +
9 c (-9+4
c)
D{\bf T} \overline{D} {\bf W}^{(\frac{3}{2})}_{0}
  - \frac{81}{10} (-9+4 c)
            \pa {\bf T}{\bf W}^{(\frac{3}{2})}_{0} \Bigg],
\nonu \\
      {\bf Q}_{5} &=& \frac{1}{(-1+c) (6+c) (-3+2 c) (-39+14 c+c^2)
        C^{(3)}_{(\frac{3}{2})(\frac{3}{2})}} \nonu \\
      & \times & \Bigg[
        \frac{9}{10} (-99+672 c-491 c^2+22 c^3+36 c^4)
              \pa^2 D  {\bf W}^{(\frac{3}{2})}_{0}
\nonu \\
&+& 3 (432-729 c+297 c^2+4 c^3+2 c^4)
      {\bf T} \pa D  {\bf W}^{(\frac{3}{2})}_{0}
\nonu \\
&+& \frac{9}{2} (-243+423 c-42 c^2+20 c^3)
            {\bf T} {\bf T} D  {\bf W}^{(\frac{3}{2})}_{0}
 +  9 (18-51 c+35 c^2)
       {\bf T} D{\bf T} {\bf W}^{(\frac{3}{2})}_{0}
 \nonu \\
 &+& \frac{3}{2} (-243+837 c-432 c^2-24 c^3+32 c^4)
      [D, \overline{D}] {\bf T} D  {\bf W}^{(\frac{3}{2})}_{0}
 \nonu \\
 &+& \frac{9}{2} (54-216 c+123 c^2-17 c^3+6 c^4)
       D {\bf T} [D, \overline{D}]  {\bf W}^{(\frac{3}{2})}_{0}
 \nonu \\
 &-& \frac{3}{2} (-702+576 c+249 c^2-193 c^3+4 c^4)
            D  {\bf T} \pa {\bf W}^{(\frac{3}{2})}_{0}
 \nonu \\
 & + & \frac{3}{2} (-918+1152 c-69 c^2-263 c^3+4 c^4)
            \pa  D
{\bf T} {\bf W}^{(\frac{3}{2})}_{0}
 \nonu \\
& - &
 \frac{3}{2} (351+81 c+108 c^2-134 c^3+8 c^4)
            \pa {\bf T} D  {\bf
W}^{(\frac{3}{2})}_{0},
\nonu \\
      {\bf Q}_{6} &=&\frac{1}{(-1+c) (6+c) (-3+2 c) (-39+14 c+c^2)
        C^{(3)}_{(\frac{3}{2})(\frac{3}{2})}}\nonu \\
      & \times & \Bigg[-\frac{
          9}{10} (-99+672 c-491 c^2+22 c^3+36 c^4)
           \pa^2 \overline{D}  {\bf
W}^{(\frac{3}{2})}_{0}
 \nonu \\
 &+ & 3 (432-729 c+297 c^2+4 c^3+2 c^4)
        {\bf T} \pa \overline{D} {\bf
W}^{(\frac{3}{2})}_{0}
 \nonu \\
 & - & \frac{9}{2} (-243+423 c-42 c^2+20 c^3)
       {\bf T} {\bf T} \overline{D}  {\bf
W}^{(\frac{3}{2})}_{0}
 -  9 (18-51 c+35 c^2)
       {\bf
T}\overline{D} {\bf T} {\bf W}^{(\frac{3}{2})}_{0}
 \nonu \\
 & - & \frac{9}{2} (54-216 c+123 c^2-17 c^3+6 c^4)
         \overline{D} {\bf T} [D, \overline{D}]  {\bf
W}^{(\frac{3}{2})}_{0}
 \nonu \\
 & - &  \frac{3}{2} (-702+576 c+249 c^2-193 c^3+4 c^4)
         \overline{D}  {\bf T} \pa {\bf
W}^{(\frac{3}{2})}_{0}
 \nonu \\
 & + & \frac{3}{2} (-918+1152 c-69 c^2-263 c^3+4 c^4)
          \pa
\overline{D} {\bf T} {\bf W}^{(\frac{3}{2})}_{0}
 \nonu \\
 & - &\frac{3}{2} (-243+837 c-432 c^2-24 c^3+32 c^4)
           [D, \overline{D}] {\bf T} \overline{D} {\bf
W}^{(\frac{3}{2})}_{0}
 \nonu \\
 & - & \frac{3}{2} (351+81 c+108 c^2-134 c^3+8 c^4)
      \pa {\bf T} \overline{D}  {\bf W}^{(\frac{3}{2})}_{0} \Bigg],
\nonu \\
      {\bf Q}_{7} &=&\frac{1}{ (-1+c) (6+c) (-3+2 c) C^{(3)}_{(\frac{3}{2})(\frac{3}{2})}} \Bigg[ \frac{9}{14} (45+6 c+8 c^2)
        {\bf T} \pa [D, \overline{D}]
{\bf W}^{(\frac{3}{2})}_{0}
 \nonu \\
 &+ & \frac{9}{7} (15+52 c)
       {\bf T} {\bf T} \pa {\bf W}^{(\frac{3}{2})}_{0}
 +  9(6+c) {\bf T}  \overline{D} {\bf T} D {\bf
W}^{(\frac{3}{2})}_{0}
+ 9(6+c) {\bf T}  D {\bf T}
\overline{D} {\bf W}^{(\frac{3}{2})}_{0}
\nonu \\
& + & \frac{9}{14} c (-93+34 c)
      \overline{D} {\bf T} \pa D {\bf
W}^{(\frac{3}{2})}_{0}
 + \frac{6}{7} (93-34 c) c
      \pa \overline{D}   {\bf T} D {\bf W}^{(\frac{3}{2})}_{0}
 \nonu \\
 & + & \frac{3}{7} (63-42 c+16 c^2)
      [D, \overline{D}] {\bf T} \pa {\bf
W}^{(\frac{3}{2})}_{0}
 -  \frac{9}{28} (63-42 c+16 c^2)
      \pa [D, \overline{D}]{\bf T}  {\bf W}^{(\frac{3}{2})}_{0}
\nonu \\
& + & \frac{9}{14} (93-34 c) c
      D {\bf T}\pa \overline{D} {\bf
W}^{(\frac{3}{2})}_{0}
+   \frac{6}{7} c (-93+34 c)
      \pa D {\bf T}\overline{D} {\bf W}^{(\frac{3}{2})}_{0}
\nonu \\
&-& \frac{9}{28} (57+2 c+40 c^2)
      \pa {\bf T} [D, \overline{D}] {\bf W}^{(\frac{3}{2})}_{0}
-  
\frac{27}{14} (15+52 c)
      \pa {\bf T} {\bf T}  {\bf W}^{(\frac{3}{2})}_{0}
\nonu \\
& + & \frac{9}{35} (42-59 c+22 c^2)
      \pa^3 {\bf W}^{(\frac{3}{2})}_{0} \Bigg].
\label{Q}
\eea
Component for (\ref{secondOPE}) can be derived using super derivatives of both sides of (\ref{secondOPE}) and applying $\theta_{12}=0$ and/or $\bar{\theta}_{12}=0$ constraints for the final stage. For the quasi primary fields components we use (\ref{stressN2}) and (\ref{supermultiplets}). The third order pole in OPEs between the stress energy tensor and   (\ref{Q}) components for $\theta=\bar{\theta}=0$ vanish as usual\footnote{ Or in ${\cal N}=2$ superspace, which is equivalent to the singular term of $\frac{\theta_{1} \bar{\theta}_{12}}{z_{12}^3}$ in OPE ${\bf T}(Z_1)\, {\bf Q}_i(Z_2)$, where $i=1,2, \cdots, 7$ vanishes. These have been explicitly checked (not shown here).}.

\section{ Operator product expansions between ${\cal N}=2$   higher spin multiplets in (\ref{supermultiplets})  }

\subsection{Operator product expansions between the first and third ${\cal N}=2$ higher spin   multiplets }

Operator product expansions between the lowest and third ${\cal N}=2$ higher spin   multiplets can be expressed as 
\bea
{\bf W}^{(\frac{3}{2})}_{0}(Z_1) \, {\bf
W}^{(\frac{7}{2})}_{0}(Z_2) & = & \frac{\theta_{12} \bar{\theta}_{12}}{z_{12}^3}
\,\,\Bigg[\frac{3 (108-144 c+15 c^2+7 c^3)}{(1-c) (-39+14 c+c^2)
    C^{(\frac{7}{2})}_{(\frac{3}{2})(3)}} \Bigg]{\bf W}^{(3)}_{0}(Z_2)
\nonu \\
& + & \frac{\theta_{12}}{z_{12}^2} \,
 \Bigg[-\frac{(432-990 c+645 c^2-79 c^3-14 c^4)}{(-1+c) (-3+2 c) (-39+14 c+c^2)C^{(\frac{7}{2})}_{(\frac{3}{2})(3)}}
 D {\bf W}^{(3)}_{0}
   + {\bf \hat{Q}}_{1} \Bigg](Z_2)
 \nonu \\
 & + & \frac{ \bar{\theta}_{12}}{z_{12}^2} \,
 \Bigg[\frac{(432-990 c+645 c^2-79 c^3-14 c^4)}{(-1+c) (-3+2 c) (-39+14 c+c^2)C^{(\frac{7}{2})}_{(\frac{3}{2})(3)}}
\overline{D} {\bf W}^{(3)}_{0}  + {\bf \hat{Q}}_{2} \Bigg](Z_2)
 \nonu \\
 & + &
\frac{\theta_{12} \bar{\theta}_{12}}{z_{12}^2} \Bigg[ \frac{1}{3}
 \pa (\mbox{pole-3})   + {\bf \hat{Q}}_{3}\Bigg](Z_2)
  \nonu \\
 & + &
\frac{1}{z_{12}} \, 
  \Bigg[\frac{c }{(-1+c) C^{(\frac{7}{2})}_{(\frac{3}{2})(3)}}
    [D, \overline{D}] {\bf
      W}^{(3)}_{0}  +{\bf \hat{Q}}_{4} \Bigg](Z_2)
\nonu \\
& + & 
\frac{\theta_{12}}{z_{12}}
\Bigg[ \frac{2}{7} \pa (\mbox{pole-2})_{\bar{\theta}=0}
+ {\bf \hat{Q}}_{5} \Bigg](Z_2)
+
\frac{\bar{\theta}_{12}}{z_{12}}
\Bigg[ \frac{2}{7} \pa (\mbox{pole-2})_{\theta=0}
+ {\bf \hat{Q}}_{6} \Bigg](Z_2)
\nonu \\
& + & \frac{\theta_{12} \bar{\theta}_{12}}{z_{12}} \,
 \Bigg[  C^{(5)}_{(\frac{3}{2})(\frac{7}{2})} {\bf W}^{(5)}_{0} 
   +\frac{1}{14} \pa^2 (\mbox{pole-3})  +\frac{3}{8} \pa
   {\bf \hat{Q}}_{3}
+ {\bf \hat{Q}}_{7} \Bigg](Z_2) +\cdots.
 \nonu
\eea
Quasi primary fields ${\bf \hat{Q}}_i(Z_2)$, which depends on ${\bf T}(Z_2); {\bf W}^{(\frac{3}{2})}_{0}(Z_2)$; and $ {\bf W}^{(3)}_{0}(Z_2)$ appear in the right hand side of the OPE, but are not explicitly expressed here for space considerations. In particular, the complete expression for ${\bf \hat{Q}}_7(Z_2)$ was not determined because we do not use OPEs between the first three ${\cal N}=2$ higher spin multiplets and the fourth (${\bf W}^{(5)}_{0}(Z_2)$). 

\subsection{ Operator product expansions between the second ${\cal N}=2$ higher spin   multiplet }

Operator product expansions between the second ${\cal N}=2$ higher spin   multiplet can be expressed as 
\bea
&& {\bf W}^{(3)}_{0}(Z_1) \, {\bf
W}^{(3)}_{0}(Z_2)  =  \frac{1}{z_{12}^6}
\,\,\Bigg[\frac{c}{3}+  3\,\theta_{12} \bar{\theta}_{12} {\bf T}\Bigg](Z_2)
\nonu \\
&& + \frac{1}{z_{12}^5} \Bigg[ -3\theta_{12} D {\bf T}
+3\bar{\theta}_{12}{\bf T} +\theta_{12} \bar{\theta}_{12}\pa \mbox{(pole-6)}\Bigg](Z_2)
+  \frac{1}{z_{12}^4} \,\frac{1}{(1-c)}\Bigg[
c [D, \overline{D}]{\bf T}
+3 {\bf T} {\bf T}
 \Bigg](Z_2)
\nonu \\
&&+  \frac{1}{z_{12}^4} \Bigg[ \theta_{12} \Bigg(
 \frac{2}{3}  \pa (\mbox{pole-5})_{\bar{\theta}=0} +  {\bf \tilde{Q}}_1 \Bigg)
  + \bar{\theta}_{12} \Bigg( \frac{2}{3}  \pa
  (\mbox{pole-5})_{\theta=0} +{\bf \tilde{Q}}_2 \Bigg)
 \Bigg](Z_2)
 \nonu \\
 &&+  \frac{\theta_{12} \bar{\theta}_{12}}{z_{12}^4}
 \Bigg[ \frac{1}{2}\pa^2 (\mbox{pole-6})+
   C_{(3)(3)}^{(3)}{\bf W}^{(3)}_{0}
 + {\bf \tilde{Q}}_3
 \Bigg](Z_2)
+ \frac{1}{z_{12}^3} \,\Bigg[
\frac{1}{2} \pa (\mbox{pole-4})_{\theta= \bar{\theta}=0}
 \Bigg](Z_2)
  \nonu \\
  &&+  \frac{\theta_{12}}{z_{12}^3} \Bigg[
    \frac{1}{4}  \pa^2 (\mbox{pole-5})_{\bar{\theta}=0}
    + \frac{3}{5} \pa  {\bf \tilde{Q}}_1 
    -\frac{(56 c^3-147 c^2+63 c-54)}{6 (c-3) c (28 c+3)}
 C_{(3)(3)}^{(3)}
     D{\bf W}^{(3)}_{0}
+ {\bf \tilde{Q}}_4
 \Bigg](Z_2)
 \nonu \\
 &&+  \frac{\bar{\theta}_{12}}{z_{12}^3} \Bigg[
   \frac{1}{4}  \pa^2 (\mbox{pole-5})_{\theta=0}+ \frac{3}{5} \pa
        {\bf \tilde{Q}}_2
   +
   \frac{(56 c^3-147 c^2+63 c-54)}{6 (c-3) c (28 c+3)}
 C_{(3)(3)}^{(3)}
\overline{D} {\bf W}^{(3)}_{0}
+ {\bf \tilde{Q}}_5
 \Bigg](Z_2)
 \nonu \\
 &&+  \frac{\theta_{12} \bar{\theta}_{12}}{z_{12}^3}
 \Bigg[\frac{1}{6}\pa^3 (\mbox{pole-6})+\frac{2}{3}
\pa \Bigg( C_{(3)(3)}^{(3)}
    {\bf W}^{(3)}_{0}
 +  {\bf \tilde{Q}}_3 \Bigg) + {\bf \tilde{Q}}_6 
 \Bigg](Z_2)
 \nonu \\
 &&+ \frac{1}{z_{12}^2} \,\Bigg[
   \frac{3}{20} \pa^2 (\mbox{pole-4})_{\theta= \bar{\theta}=0}
-\frac{(8 c^2+9)}{6 (c-3) (28 c+3)} C_{(3)(3)}^{(3)}
   [D, \overline{D}] {\bf W}^{(3)}_{0}
+ {\bf \tilde{Q}}_7 
 \Bigg](Z_2)
  \nonu \\
  &&+  \frac{\theta_{12}}{z_{12}^2} \Bigg[
    \frac{1}{15}  \pa^3 (\mbox{pole-5})_{\bar{\theta}=0}
    + \frac{1}{5} \pa^2  {\bf \tilde{Q}}_1
     +
     \frac{4}{7} \pa \Bigg(
     \mbox{last two terms in}
 \frac{\theta_{12}}{z_{12}^3}
 \Bigg)
    +{\bf \tilde{Q}}_8
 \Bigg](Z_2)
 \nonu \\
 &&+  \frac{\bar{\theta}_{12}}{z_{12}^2} \Bigg[
    \frac{1}{15}  \pa^3 (\mbox{pole-5})_{\theta=0}
+ \frac{1}{5} \pa^2  {\bf \tilde{Q}}_2
   + \frac{4}{7} \pa \Bigg(
 \mbox{last two terms in}
 \frac{\bar{\theta}_{12}}{z_{12}^3}
   \Bigg)
 +
   {\bf \tilde{Q}}_9 
 \Bigg](Z_2)
 \nonu \\
 &&+  \frac{\theta_{12} \bar{\theta}_{12}}{z_{12}^2}
 \Bigg[
   \frac{1}{24}\pa^4 (\mbox{pole-6})
   +\frac{5}{21} \pa^2 \Bigg(
C_{(3)(3)}^{(3)}
   {\bf W}^{(3)}_{0}
 +  {\bf \tilde{Q}}_3 \Bigg)
+ 
  \frac{5}{8} \pa {\bf \tilde{Q}}_6
  \nonu \\
  &&
  -\frac{5 C^{(5)}_{(\frac{3}{2}){(\frac{7}{2})}}
    C^{(\frac{7}{2})}_{(\frac{3}{2}){(\frac{3}{2})}}}{C^{(3)}_{(\frac{3}{2}){(\frac{3}{2})}}}
 {\bf W}^{(5)}_{0}
 + {\bf \tilde{Q}}_{10} 
 \Bigg](Z_2)
\nonu \\
&& + 
\frac{1}{z_{12}} \,\Bigg[
  \frac{1}{30} \pa^3 (\mbox{pole-4})_{\theta= \bar{\theta}=0}
  + \frac{1}{2} \pa \Bigg(
\mbox{last two terms in}
 \frac{1}{z_{12}^2}
\Bigg)
  \Bigg](Z_2)
  \nonu \\
  &&+  \frac{\theta_{12}}{z_{12}} \Bigg[
    \frac{1}{72}  \pa^4 (\mbox{pole-5})_{\bar{\theta}=0}
+ \frac{1}{21} \pa^3  {\bf \tilde{Q}}_1
   +\frac{5}{28}
   \pa^2 \Bigg(
\mbox{last two terms in}
 \frac{\theta_{12}}{z_{12}^3}
   \Bigg) \nonu \\
  &&+\frac{5}{9} \pa {\bf \tilde{Q}}_8  +
   \frac{C^{(5)}_{(\frac{3}{2}){(\frac{7}{2})}}
     C^{(\frac{7}{2})}_{(\frac{3}{2}){(\frac{3}{2})}}}
        {C^{(3)}_{(\frac{3}{2}){(\frac{3}{2})}}}
  D {\bf W}^{(5)}_{0}  +{\bf \tilde{Q}}_{11}
 \Bigg](Z_2)
 \nonu \\
 &&+
 \frac{\bar{\theta}_{12}}{z_{12}} \Bigg[
 \frac{1}{72}  \pa^4 (\mbox{pole-5})_{\theta=0}
 + \frac{1}{21} \pa^3  {\bf \tilde{Q}}_2
 + \frac{5}{28} \pa^2 \Bigg(
\mbox{last two terms in}
 \frac{\bar{\theta}_{12}}{z_{12}^3} 
 \Bigg)
 \nonu \\
 &&
 +\frac{5}{9} \pa {\bf \tilde{Q}}_9 
 -\frac{C^{(5)}_{(\frac{3}{2}){(\frac{7}{2})}}
   C^{(\frac{7}{2})}_{(\frac{3}{2}){(\frac{3}{2})}}}{C^{(3)}_{(\frac{3}{2}){(\frac{3}{2})}}}
  \overline{D} {\bf W}^{(5)}_{0}  +{\bf \tilde{Q}}_{12}
\Bigg](Z_2)
 \nonu \\
 &&+  \frac{\theta_{12} \bar{\theta}_{12}}{z_{12}}
 \Bigg[
   \frac{1}{120}\pa^5 (\mbox{pole-6})
+ \frac{5}{84} \pa^3 \Bigg(C_{(3)(3)}^{(3)}
 {\bf W}^{(3)}_{0} +{\bf \tilde{Q}}_3 \Bigg)
+ 
\frac{5}{24} \pa^2 {\bf \tilde{Q}}_6
\nonu \\
&&  +\frac{3}{5} \pa \Bigg(-\frac{ 5
  C^{(5)}_{(\frac{3}{2}){(\frac{7}{2})}}
  C^{(\frac{7}{2})}_{(\frac{3}{2}){(\frac{3}{2})}}}{C^{(3)}_{(\frac{3}{2}){(\frac{3}{2})}}}
 {\bf W}^{(5)}_{0}
 +  {\bf \tilde{Q}}_{10} \Bigg) +
   {\bf \tilde{Q}}_{13}
 \Bigg](Z_2)
 + \cdots,
\nonu 
\eea
where the self-coupling constant 
\bea
(C_{(3)(3)}^{(3)})^2 =
\frac{24 (c-3)^2 c^2 (28 c+3)^2}{
  (c-1) (c+6) (2 c-3) (2 c+3) (4 c-9) (5 c-3)},
\nonu
\eea
and we omit the quasi primary fields ${\bf \tilde{Q}}_i(Z_2)$ $(i=1, 2, \cdots, 13)$, which depends on ${\bf T}(Z_2); {\bf W}^{(\frac{3}{2})}_{0}(Z_2)$; and $ {\bf W}^{(3)}_{0}(Z_2)$. As above, the complete expressions for ${\bf \tilde{Q}}_{10}(Z_2)$, ${\bf \tilde{Q}}_{11}(Z_2)$, ${\bf \tilde{Q}}_{12}(Z_2)$, and ${\bf \tilde{Q}}_{13}(Z_2)$ were not determined.


\subsection{Operator product expansions between the second and the third ${\cal N}=2$ higher spin   multiplets }

Operator product expansions between the second and the third ${\cal N}=2$ higher spin   multiplets can be expressed as 
\bea
&&{\bf W}^{(3)}_{0}(Z_1) \, {\bf
W}^{(\frac{7}{2})}_{0}(Z_2) = \frac{\theta_{12} \bar{\theta}_{12} }{z_{12}^6}
\,\,\Bigg[-\frac{3 (6 + c) (18 - 27 c + 7 c^2) }{2 (-1 + c) (-39 + 14 c + c^2) C^{(\frac{7}{2})}_{(\frac{3}{2}){(\frac{3}{2})}}}
\Bigg] {\bf W}^{(\frac{3}{2})}_{0}(Z_2)
\nonu \\
&&+
\frac{1}{z_{12}^5} \frac{(6 + c) (18 - 27 c + 7 c^2) }{(-1 + c) (-39 + 14 c + c^2) C^{(\frac{7}{2})}_{(\frac{3}{2}){(\frac{3}{2})}}}\Bigg[ \theta_{12} D {\bf W}^{(\frac{3}{2})}_{0}
-\bar{\theta}_{12} \overline{D} {\bf W}^{(\frac{3}{2})}_{0}\Bigg](Z_2)
+\frac{\theta_{12} \bar{\theta}_{12}}{z_{12}^5} \Bigg[\frac{2}{3}\pa \mbox{(pole-6)}\Bigg](Z_2)
\nonu \\
&&+  \frac{1}{z_{12}^4} \,\frac{(6 + c) (18 - 27 c + 7 c^2) }{(-1 + c) (-9 + 4 c) (-39 + 14 c + c^2) C^{(\frac{7}{2})}_{(\frac{3}{2}){(\frac{3}{2})}}} \Bigg[ c [D, \overline{D}] {\bf
W}^{(\frac{3}{2})}_{0} +9 {\bf T} {\bf W}^{(\frac{3}{2})}_{0}
 \Bigg](Z_2)
\nonu \\
&&+  \frac{1}{z_{12}^4} \Bigg[ \theta_{12} \Bigg(
 \frac{1}{2}  \pa (\mbox{pole-5})_{\bar{\theta}=0} +  {\bf \check{Q}}_1 \Bigg)
  + \bar{\theta}_{12} \Bigg( \frac{1}{2}
  \pa (\mbox{pole-5})_{\theta=0} + {\bf \check{Q}}_2 \Bigg)
 \Bigg](Z_2)
 \nonu \\
 &&+  \frac{\theta_{12} \bar{\theta}_{12}}{z_{12}^4}
 \Bigg[ \frac{1}{4}\pa^2 (\mbox{pole-6})+
   \frac{21 (675 - 531 c + 3687 c^2 - 2527 c^3 + 220 c^4 + 48 c^5)}{4 (-1 + c) (6 + c) (-3 + 2 c) (-39 + 14 c + c^2)
     C^{(3)}_{(\frac{3}{2}){(\frac{3}{2})}}}{\bf W}^{(\frac{7}{2})}_{0}
 + {\bf \check{Q}}_3
 \Bigg](Z_2)
 \nonu \\
 &&+ \frac{1}{z_{12}^3} \,\Bigg[
\frac{2}{5} \pa (\mbox{pole-4})_{\theta= \bar{\theta}=0} +{\bf \check{Q}}_4
 \Bigg](Z_2)
  +  \frac{\theta_{12}}{z_{12}^3} \Bigg[
    \frac{3}{20}  \pa^2 (\mbox{pole-5})_{\bar{\theta}=0}
     + 
\frac{1}{2} \pa  {\bf {\check{Q}}}_1 
\nonu \\
&&
-\frac{3 (675 - 531 c + 3687 c^2 - 2527 c^3 + 220 c^4 + 48 c^5)}{2 (-1 + c) (6 + c) (-3 + 2 c) (-39 + 14 c + c^2) C^{(3)}_{(\frac{3}{2}){(\frac{3}{2})}}} D{\bf W}^{(\frac{7}{2})}_{0} +{\bf \check{Q}}_5 \Bigg](Z_2)
 \nonu \\
 &&+  \frac{\bar{\theta}_{12}}{z_{12}^3}\Bigg[
   \frac{3}{20}  \pa^2 (\mbox{pole-5})_{\theta=0}
    + 
    \frac{1}{2} \pa  {\bf \check{Q}}_2
    \nonu \\
    && +
    \frac{3 (675 - 531 c + 3687 c^2 - 2527 c^3 + 220 c^4 + 48 c^5)}{2 (-1 + c) (6 + c) (-3 + 2 c) (-39 + 14 c + c^2) C^{(3)}_{(\frac{3}{2}){(\frac{3}{2})}}} \overline{D}
{\bf W}^{(\frac{7}{2})}_{0}+ {\bf \check{Q}}_6 
 \Bigg](Z_2)
\nonu \\
&&+ \frac{\theta_{12} \bar{\theta}_{12}}{z_{12}^3}
 \Bigg[{\bf \check{Q}}_7 +\frac{1}{15}\pa^3 (\mbox{pole-6})
 +\frac{4}{7} \pa \Bigg( \mbox{last two terms in}
 \frac{\theta_{12} \bar{\theta}_{12}}{z_{12}^4}
 \Bigg )
 \Bigg](Z_2)
 \nonu \\
 &&+ \frac{1}{z_{12}^2} \,\Bigg[
\frac{1}{10} \pa^2 (\mbox{pole-4})_{\theta= \bar{\theta}=0}
+ \frac{3}{7} \pa {\bf \check{Q}}_4
\nonu \\
&&   -\frac{3 c (-63 - 9 c - 101 c^2 + 59 c^3 + 6 c^4)}
 {2 (-1 + c) (6 + c) (-3 + 2 c) (-39 + 14 c + c^2) C^{(3)}_{(\frac{3}{2}){(\frac{3}{2})}}}[D, \overline{D}]{\bf W}^{(\frac{7}{2})}_{0}
 + {\bf \check{Q}}_8
\Bigg](Z_2)
  \nonu \\
  &&+  \frac{\theta_{12}}{z_{12}^2} \Bigg[
    \frac{1}{30}  \pa^3 (\mbox{pole-5})_{\bar{\theta}=0}
    + 
\frac{1}{7} \pa^2  {\bf \check{Q}}_1 +
    \frac{1}{2}\pa \Bigg (
\mbox{last two terms in}
 \frac{{\theta}_{12}} {z_{12}^3}    
    \Bigg )
+{\bf \check{Q}}_9
 \Bigg](Z_2)
 \nonu \\
 &&+  \frac{\bar{\theta}_{12}}{z_{12}^2} \Bigg[
   \frac{1}{30}  \pa^3 (\mbox{pole-5})_{\theta=0}
+ 
\frac{1}{7} \pa^2  {\bf \check{Q}}_2
   +\frac{1}{2}\pa \Bigg (
 \mbox{last two terms in}
 \frac{\bar{\theta}_{12}} {z_{12}^3}
   \Bigg)
  +{\bf \check{Q}}_{10}
 \Bigg](Z_2)
 \nonu \\
 &&+  \frac{\theta_{12} \bar{\theta}_{12}}{z_{12}^2}
 \Bigg[ \frac{1}{72}\pa^4 (\mbox{pole-6})
 +\frac{5}{28} \pa^2 \Bigg(
\mbox{last two terms in}
 \frac{\theta_{12} \bar{\theta}_{12}}{z_{12}^4}
 \Bigg) +\frac{5}{9}\pa {\bf \check{Q}}_7 +
 C^{(\frac{11}{2})}_{(3) (\frac{7}{2})} {\bf W}^{(\frac{11}{2})}_{0}+ {\bf \check{Q}}_{11}
 \Bigg](Z_2)
  \nonu \\
&&+
  \frac{1}{z_{12}} \,\Bigg[\frac{2}{105} \pa^3 (\mbox{pole-4})_{\theta= \bar{\theta}=0} 
 + \frac{3}{28} \pa^2 {\bf \check{Q}}_4
    +\frac{4}{9} \pa \Bigg (
\mbox{last two terms in}
 \frac{1}{z_{12}^2}
 \Bigg )
 +{\bf \check{Q}}_{12}
 \Bigg](Z_2)
  \nonu \\
  &&+  \frac{\theta_{12}}{z_{12}} \Bigg[
    \frac{1}{168}  \pa^4 (\mbox{pole-5})_{\bar{\theta}=0}
  + 
\frac{5}{168} \pa^3  {\bf \check{Q}}_1  +\frac{5}{36}\pa^2 \Bigg (
\mbox{last two terms in}
 \frac{{\theta}_{12}} {z_{12}^3}   
    \Bigg )
+\frac{1}{2} \pa {\bf \check{Q}}_9
\nonu \\
&& -\frac{2}{11} C^{(\frac{11}{2})}_{(3) (\frac{7}{2})}
      {\bf W}^{(\frac{11}{2})}_{0}
+{\bf \check{Q}}_{13}
 \Bigg](Z_2)
 \nonu \\
 &&+  \frac{\bar{\theta}_{12}}{z_{12}} \Bigg[
   \frac{1}{168}  \pa^4 (\mbox{pole-5})_{\theta=0}
+ 
\frac{5}{168} \pa^3  {\bf \check{Q}}_2
   +
   \frac{5}{36}\pa^2 \Bigg (
\mbox{last two terms in}
 \frac{\bar{\theta}_{12}} {z_{12}^3}
   \Bigg )
  +\frac{1}{2} \pa {\bf \check{Q}}_{10}
  \nonu \\
  && 
  +\frac{2}{11} C^{(\frac{11}{2})}_{(3) (\frac{7}{2})}
  {\bf W}^{(\frac{11}{2})}_{0}
+{\bf \check{Q}}_{14}
 \Bigg](Z_2)
 \nonu \\
 && + \frac{\theta_{12} \bar{\theta}_{12}}{z_{12}}
 \Bigg[\frac{1}{420}\pa^5 (\mbox{pole-6})
+\frac{5}{126} \pa^3 \Bigg(
\mbox{last two terms in}
 \frac{\theta_{12} \bar{\theta}_{12}}{z_{12}^4}
 \Bigg ) + \frac{1}{6}\pa^2 {\bf \check{Q}}_7 
 \nonu \\
 && +\frac{6}{11} \pa \Bigg ( C^{(\frac{11}{2})}_{(3) (\frac{7}{2})}
       {\bf W}^{(\frac{11}{2})}_{0} +{\bf \check{Q}}_{11} \Bigg ) +
   {\bf \check{Q}}_{15}
 \Bigg](Z_2) + \cdots ,
\nonu
\eea
where we omit the quasi primary fields ${\bf \check{Q}}_i(Z_2)$ $(i=1, 2, \cdots, 15)$, which depends on ${\bf T}(Z_2); {\bf W}^{(\frac{3}{2})}_{0}(Z_2)$; and $ {\bf W}^{(\frac{7}{2})}_{0}(Z_2)$. Similarly to the previous derivations, complete expressions for ${\bf \check{Q}}_{11}(Z_2)$,${\bf \check{Q}}_{13}(Z_2)$, ${\bf \check{Q}}_{14}(Z_2)$, and ${\bf \check{Q}}_{15}(Z_2)$ were not determined. 

\subsection{Operator product expansions between the third ${\cal N}=2$ higher spin   multiplet }

Operator product expansions between the third ${\cal N}=2$ higher spin   multiplet can be expressed as
\bea
&&{\bf W}^{(\frac{7}{2})}_{0}(Z_1) \, {\bf
W}^{(\frac{7}{2})}_{0}(Z_2) = \frac{1 }{z_{12}^7}
\,\,\Bigg[\frac{2 c}{7} +3 \theta_{12} \bar{\theta}_{12} {\bf T}\Bigg](Z_2)
\nonu \\
&+&\frac{1 }{z_{12}^6}
\,\,\Bigg[3(- \theta_{12} D {\bf T} +\bar{\theta}_{12} \overline{D} {\bf T}) + \theta_{12} \bar{\theta}_{12} \pa (\mbox{pole-7})\Bigg](Z_2)
+ \frac{1}{z_{12}^5} \frac{1}{(1-c)}\Bigg[ c [D,\overline{D}] {\bf T} + 3 {\bf T} {\bf T}\Bigg](Z_2)
\nonu \\
&+& \frac{1}{z_{12}^5} \Bigg[ \theta_{12} \Bigg(
  \frac{2}{3}  \pa (\mbox{pole-6})_{\bar{\theta}=0} +
   {\bf \breve{Q}}_1 \Bigg)
  +\bar{\theta}_{12} \Bigg(
  \frac{2}{3} \pa (\mbox{pole-6})_{\theta=0} +
   {\bf \breve{Q}}_2 \Bigg)\Bigg](Z_2)
\nonu \\
&+& \frac{\theta_{12} \bar{\theta}_{12}}{z_{12}^5} \Bigg[
   \frac{1}{2}  \pa^2 (\mbox{pole-7})+
  C^{(3)}_{(\frac{7}{2})(\frac{7}{2})} {\bf W}^{(3)}_{0}
+ {\bf \breve{Q}}_3 \Bigg](Z_2)
+ \frac{1}{z_{12}^4} \Bigg[ \frac{1}{2}  \pa (\mbox{pole-5})_{\theta=\bar{\theta}=0}\Bigg](Z_2)
\nonu \\
&+& \frac{\theta_{12}}{z_{12}^4} \Bigg[ 
\frac{1}{4}  \pa^2 (\mbox{pole-6})_{\bar{\theta}=0}
+\frac{3}{5} \pa {\bf \breve{Q}}_1  +C^{(\frac{7}{2})}_{(\frac{7}{2})(\frac{7}{2})} D {\bf W}^{(3)}_{0} + {\bf \breve{Q}}_4
\Bigg](Z_2)
\nonu \\
&+& \frac{\bar{\theta}_{12}}{z_{12}^4} \Bigg[
 \frac{1}{4}  \pa^2 (\mbox{pole-6})_{\theta=0}
+\frac{3}{5} \pa {\bf \breve{Q}}_2 - C^{(\frac{7}{2})}_{(\frac{7}{2})(\frac{7}{2})} \overline{D} {\bf W}^{(3)}_{0} + {\bf \breve{Q}}_5
\Bigg](Z_2)
\nonu \\
&+& \frac{\theta_{12} \bar{\theta}_{12}}{z_{12}^4} \Bigg[
 \frac{1}{6}  \pa^3 (\mbox{pole-7})+\frac{2}{3} \pa \Bigg (C^{(3)}_{(\frac{7}{2})(\frac{7}{2})} {\bf W}^{(3)}_{0}
+ {\bf \breve{Q}}_3 \Bigg )+ {\bf \breve{Q}}_6
 \Bigg](Z_2)
\nonu \\
&+& \frac{1}{z_{12}^3} \Bigg[ 
\frac{3}{20}  \pa^2 (\mbox{pole-5})_{\theta=\bar{\theta}=0}+
  C^{(4)}_{(\frac{7}{2})(\frac{7}{2})}[D,\overline{D}] {\bf W}^{(3)}_{0} + {\bf \breve{Q}}_7 \Bigg](Z_2)
\nonu \\
&+& \frac{\theta_{12}}{z_{12}^3} \Bigg[ 
  \frac{1}{15}  \pa^3 (\mbox{pole-6})_{\bar{\theta}=0}
+\frac{1}{5} \pa^2 {\bf \breve{Q}}_1 
+\frac{4}{7} \pa \Bigg (C^{(\frac{7}{2})}_{(\frac{7}{2})(\frac{7}{2})} D {\bf W}^{(3)}_{0} + {\bf \breve{Q}}_4 \Bigg ) +
  {\bf \breve{Q}}_8
\Bigg](Z_2)
\nonu \\
&+& \frac{\bar{\theta}_{12}}{z_{12}^3} \Bigg[
  \frac{1}{15}  \pa^3 (\mbox{pole-6})_{\theta=0}
  +\frac{1}{5} \pa^2 {\bf \breve{Q}}_2
+\frac{4}{7} \pa \Bigg (- C^{(\frac{7}{2})}_{(\frac{7}{2})(\frac{7}{2})} \overline{D} {\bf W}^{(3)}_{0} + {\bf \breve{Q}}_5 \Bigg )
+{\bf \breve{Q}}_9
\Bigg](Z_2)
\nonu \\
&+& \frac{\theta_{12} \bar{\theta}_{12}}{z_{12}^3} \Bigg[
\frac{1}{24}  \pa^4
(\mbox{pole-7})
+\frac{5}{21} \pa^2 \Bigg (C^{(3)}_{(\frac{7}{2})(\frac{7}{2})} {\bf W}^{(3)}_{0}
+ {\bf \breve{Q}}_3 \Bigg )
+\frac{5}{8} \pa {\bf \breve{Q}}_6
+ 
C^{(5)}_{(\frac{7}{2})(\frac{7}{2})} {\bf W}^{(5)}_{0} +{\bf \breve{Q}}_{10}
\Bigg](Z_2)
\nonu \\
&+& \frac{1}{z_{12}^2} \Bigg[ 
\frac{1}{30}  \pa^3 (\mbox{pole-5})_{\theta=\bar{\theta}=0}+
 \frac{1}{2} \pa \Bigg (
C^{(4)}_{(\frac{7}{2})(\frac{7}{2})}[D,\overline{D}] {\bf W}^{(3)}_{0} + {\bf \breve{Q}}_7 \Bigg )
\Bigg](Z_2)
\nonu \\
&+& \frac{\theta_{12}}{z_{12}^2} \Bigg[
  \frac{1}{72}  \pa^4 (\mbox{pole-6})_{\bar{\theta}=0}
  +\frac{1}{21} \pa^3 {\bf \breve{Q}}_1 
+\frac{5}{28} \pa^2 \Bigg (C^{(\frac{7}{2})}_{(\frac{7}{2})(\frac{7}{2})} D {\bf W}^{(3)}_{0}
  + {\bf \breve{Q}}_4 \Bigg )
  +\frac{5}{9} \pa {\bf \breve{Q}}_8
\nonu \\
&+&
C^{(\frac{11}{2})}_{(\frac{7}{2})(\frac{7}{2})} D {\bf W}^{(5)}_{0} + {\bf \breve{Q}}_{11}
\Bigg](Z_2)
\nonu \\
&+& \frac{\bar{\theta}_{12}}{z_{12}^2} \Bigg[
\frac{1}{72}  \pa^4 (\mbox{pole-6})_{\theta=0}
+\frac{1}{21} \pa^3 {\bf \breve{Q}}_2
+\frac{5}{28} \pa^2 \Bigg (-C^{(\frac{7}{2})}_{(\frac{7}{2})(\frac{7}{2})} \overline{D }  {\bf W}^{(3)}_{0} + {\bf \breve{Q}}_5 \Bigg )
+\frac{5}{9} \pa {\bf \breve{Q}}_9
\nonu \\
&-&
C^{(\frac{11}{2})}_{(\frac{7}{2})(\frac{7}{2})} \overline{D } {\bf W}^{(5)}_{0} + {\bf \breve{Q}}_{12}
\Bigg](Z_2)
\nonu \\
&+& \frac{\theta_{12} \bar{\theta}_{12}}{z_{12}^2} \Bigg[
  \frac{1}{120}  \pa^5 (\mbox{pole-7})
  +\frac{5}{84} \pa^3 \Bigg (C^{(3)}_{(\frac{7}{2})(\frac{7}{2})} {\bf W}^{(3)}_{0} 
  +{\bf \breve{Q}}_3 \Bigg )
  + \frac{5}{24} \pa^2 {\bf \breve{Q}}_6
+\frac{3}{5} \pa \Bigg (C^{(5)}_{(\frac{7}{2})(\frac{7}{2})} {\bf W}^{(5)}_{0} +{\bf \breve{Q}}_{10} \Bigg )
\nonu \\
&+&  
{\bf \breve{Q}}_{13}
\Bigg](Z_2)
\nonu \\
&+& \frac{1}{z_{12}} \Bigg[
\frac{1}{168}  \pa^4 (\mbox{pole-5})_{\theta=\bar{\theta}=0}
+\frac{5}{36} \pa^2 \Bigg ( C^{(4)}_{(\frac{7}{2})(\frac{7}{2})}[D,\overline{D}] {\bf W}^{(3)}_{0} + {\bf \breve{Q}}_7 \Bigg )
+C^{(6)}_{(\frac{7}{2})(\frac{7}{2})} [D,\overline{D}]{\bf W}^{(5)}_{0}
 +{\bf \breve{Q}}_{14}
\Bigg](Z_2)
\nonu \\
&+& \frac{\theta_{12}}{z_{12}} \Bigg[
   \frac{1}{420}  \pa^5 (\mbox{pole-6})_{\bar{\theta}=0}
+\frac{1}{112} \pa^4 {\bf \breve{Q}}_1
+\frac{5}{126} \pa^3 \Bigg (C^{(\frac{7}{2})}_{(\frac{7}{2})(\frac{7}{2})} D {\bf W}^{(3)}_{0} + {\bf \breve{Q}}_4 \Bigg )
 +\frac{1}{6} \pa^2 {\bf \breve{Q}}_8
\nonu \\
&+&
 \frac{6}{11} \pa \Bigg (C^{(\frac{11}{2})}_{(\frac{7}{2})(\frac{7}{2})}
D {\bf W}^{(5)}_{0}
 + {\bf \breve{Q}}_{11} \Bigg )
+{\bf \breve{Q}}_{15} 
\Bigg](Z_2)
\nonu \\
&+& \frac{\bar{\theta}_{12}}{z_{12}} \Bigg[
   \frac{1}{420}  \pa^5 (\mbox{pole-6})_{\theta=0}+
  \frac{1}{112} \pa^4 {\bf \breve{Q}}_2
+\frac{5}{126} \pa^3 \Bigg (-C^{(\frac{7}{2})}_{(\frac{7}{2})(\frac{7}{2})} \overline{D }  {\bf W}^{(3)}_{0} + {\bf \breve{Q}}_5 \Bigg )
+\frac{1}{6} \pa^2 {\bf \breve{Q}}_9
\nonu \\
&+&
\frac{6}{11} \pa \Bigg (-C^{(\frac{11}{2})}_{(\frac{7}{2})(\frac{7}{2})} \overline{D } {\bf W}^{(5)}_{0} + {\bf \breve{Q}}_{12} \Bigg )
+ {\bf \breve{Q}}_{16} 
\Bigg](Z_2)
\nonu \\
&+& \frac{\theta_{12} \bar{\theta}_{12}}{z_{12}} \Bigg[
   \frac{1}{720}  \pa^6 (\mbox{pole-7})
  +\frac{5}{432} \pa^4 \Bigg (C^{(3)}_{(\frac{7}{2})(\frac{7}{2})} {\bf W}^{(3)}_{0}
  +{\bf \breve{Q}}_3 \Bigg )
  +\frac{7}{144} \pa^3 {\bf \breve{Q}}_6
\nonu \\
&+&\frac{21}{110} \pa^2 \Bigg (C^{(5)}_{(\frac{7}{2})(\frac{7}{2})} {\bf W}^{(5)}_{0} +{\bf \breve{Q}}_{10} \Bigg )+
\frac{7}{12} \pa {\bf \breve{Q}}_{13}
+C^{(7)}_{(\frac{7}{2})(\frac{7}{2})} {\bf W}^{(7)}_{0} +{\bf \breve{Q}}_{17}
\Bigg](Z_2)
+ \cdots.
\nonu
\eea
where the structure constants  are given by
\bea
(C^{(3)}_{(\frac{7}{2})(\frac{7}{2})})^{2} &=& \frac{27 (675 - 531 c + 3687 c^2 - 2527 c^3 + 220 c^4 + 48 c^5)^2}{2 (-1 + c) (6 + c) (-3 + 2 c) (3 + 2 c) (-9 + 4 c) (-3 + 5 c) (-39 + 
   14 c + c^2)^2},
\nonu \\
(C^{(\frac{7}{2})}_{(\frac{7}{2})(\frac{7}{2})})^{2}&=& \frac{6 (-3942 + 9369 c - 12222 c^2 + 9646 c^3 - 3371 c^4 + 238 c^5 + 
   48 c^6)^2}{(-1 + c) (6 + c) (-3 + 2 c)^3 (3 + 2 c) (-9 + 4 c) (-3 + 5 c) (-39 + 
   14 c + c^2)^2},
\nonu \\
(C^{(4)}_{(\frac{7}{2})(\frac{7}{2})})^{2}& =& 
\frac{54 c^2 (4968 - 14202 c + 19152 c^2 - 15215 c^3 + 7000 c^4 - 
   1593 c^5 + 78 c^6 + 16 c^7)^2}{(-1 + c) (6 + c) (-3 + 2 c)^3 (3 + 2 c) (-9 + 4 c) (-3 + 5 c) }
\nonu \\
&\times& \frac{1}{ (-39 +    14 c + c^2)^2 (18 - 27 c + 7 c^2)^2},
\nonu \\
(C^{(5)}_{(\frac{7}{2})(\frac{7}{2})})^{2} &=& \frac{50 (216 - 459 c + 114 c^2 + 91 c^3 - 22 c^4)^2 (C^{(5)}_{(\frac{3}{2})(\frac{7}{2})})^2}{3 (-3 + 2 c) (3 + 2 c) (-9 + 4 c) (-3 + 5 c) (-39 + 14 c + c^2) (18 - 
   27 c + 7 c^2)},
\nonu \\
(C^{(\frac{11}{2})}_{(\frac{7}{2})(\frac{7}{2})})^{2} &=& \frac{2 (-513 + 1269 c - 855 c^2 + 89 c^3 + 22 c^4)^2 (C^{(5)}_{(\frac{3}{2})(\frac{7}{2})})^2}{3 (-3 + 2 c) (3 + 2 c) (-9 + 4 c) (-3 + 5 c) (-39 + 14 c + c^2) (18 - 
   27 c + 7 c^2)},
\nonu \\
(C^{(6)}_{(\frac{7}{2})(\frac{7}{2})})^{2} &=& \frac{2 c^2 (-3 + 2 c) (-39 + 14 c + c^2) (C^{(5)}_{(\frac{3}{2})(\frac{7}{2})})^2}{3 (3 + 2 c) (-9 + 4 c) (-3 + 5 c) (18 - 27 c + 7 c^2)}.
\nonu 
\eea
$C^{(7)}_{(\frac{7}{2})(\frac{7}{2})} $ is undetermined.
The quasi primary fields ${\bf \breve{Q}}_{i}$ ($i=11, 12, \cdots, 17$), remain undetermined.

The analysis from Section $5$ can be similarly described.

\section{ Partners for (\ref{nextlowest}) from Section $5.2$ }

The remaining (\ref{nextlowest}) generators for the second ${\cal N}=2$ higher spin multiplet can be obtained from the relationships in footnote~\ref{remain}, 
\bea
&&V^{(\frac{7}{2})+}_{\frac{5}{2}} = 2(\frac{-i}{4})^{\frac{5}{2}}
\hat{y}_1 \hat{y}_1 \hat{y}_1 \hat{y}_1 \hat{y}_1 \otimes 
  \left( {\begin{array}{cc}
  0 & 1 \\
   0 & 0 \\
  \end{array} } \right), \,
\nonu \\ 
&&V^{(\frac{7}{2})+}_{\frac{3}{2}} = (\frac{-i}{4})^{\frac{5}{2}}
\frac{2}{5}
(\hat{y}_1 \hat{y}_1 \hat{y}_1 \hat{y}_1 \hat{y}_2 +
\hat{y}_1 \hat{y}_1 \hat{y}_1 \hat{y}_2 \hat{y}_1+
\hat{y}_1 \hat{y}_1 \hat{y}_2 \hat{y}_1 \hat{y}_1 +
\hat{y}_1 \hat{y}_2 \hat{y}_1 \hat{y}_1 \hat{y}_1+
\hat{y}_2 \hat{y}_1 \hat{y}_1 \hat{y}_1 \hat{y}_1 
) 
  \otimes \left( {\begin{array}{cc}
    0 & 1 \\
   0 & 0 \\
  \end{array} } \right), 
\nonu \\ 
&&V^{(\frac{7}{2})+}_{\frac{1}{2}} = (\frac{-i}{4})^{\frac{5}{2}}
\frac{2}{10} (\hat{y}_1 \hat{y}_1 \hat{y}_1 \hat{y}_2 \hat{y}_2 +
\hat{y}_1 \hat{y}_1 \hat{y}_2\hat{y}_1 \hat{y}_2 +
\hat{y}_1 \hat{y}_1 \hat{y}_2 \hat{y}_2 \hat{y}_1+
  \hat{y}_1 \hat{y}_2 \hat{y}_1 \hat{y}_1 \hat{y}_2 
   +
   \hat{y}_1 \hat{y}_2 \hat{y}_1\hat{y}_2 \hat{y}_1
   \nonu \\
   && + \hat{y}_1 \hat{y}_2\hat{y}_2 \hat{y}_1 \hat{y}_1
   +\hat{y}_2 \hat{y}_1 \hat{y}_1 \hat{y}_1 \hat{y}_2 +
   \hat{y}_2 \hat{y}_1 \hat{y}_1\hat{y}_2 \hat{y}_1 +
   \hat{y}_2 \hat{y}_1\hat{y}_2 \hat{y}_1 \hat{y}_1+
  \hat{y}_2 \hat{y}_2 \hat{y}_1 \hat{y}_1 \hat{y}_1 
   ) 
  \otimes \left( {\begin{array}{cc}
    0 & 1 \\
   0 & 0 \\
  \end{array} } \right), \,
  \nonu \\
&&V^{(\frac{7}{2})+}_{-\frac{1}{2}} = (\frac{-i}{4})^{\frac{5}{2}}
\frac{2}{10} (\hat{y}_2 \hat{y}_2 \hat{y}_2 \hat{y}_1 \hat{y}_1 +
\hat{y}_2 \hat{y}_2 \hat{y}_1\hat{y}_2 \hat{y}_1 +
\hat{y}_2 \hat{y}_2 \hat{y}_1 \hat{y}_1 \hat{y}_2+
  \hat{y}_2 \hat{y}_1 \hat{y}_2 \hat{y}_2 \hat{y}_1 
   +
   \hat{y}_2 \hat{y}_1 \hat{y}_2\hat{y}_1 \hat{y}_2
   \nonu \\
   && + \hat{y}_2 \hat{y}_1\hat{y}_1 \hat{y}_2 \hat{y}_2
   +\hat{y}_1 \hat{y}_2 \hat{y}_2 \hat{y}_2 \hat{y}_1 +
   \hat{y}_1 \hat{y}_2 \hat{y}_2\hat{y}_1 \hat{y}_2 +
   \hat{y}_1 \hat{y}_2\hat{y}_1 \hat{y}_2 \hat{y}_2+
  \hat{y}_1 \hat{y}_1 \hat{y}_2 \hat{y}_2 \hat{y}_2 
   ) 
  \otimes \left( {\begin{array}{cc}
    0 & 1 \\
   0 & 0 \\
  \end{array} } \right), \,
  \nonu \\
  &&V^{(\frac{7}{2})+}_{-\frac{3}{2}} = (\frac{-i}{4})^{\frac{5}{2}}
\frac{2}{5}
(\hat{y}_2 \hat{y}_2 \hat{y}_2 \hat{y}_2 \hat{y}_1 +
\hat{y}_2 \hat{y}_2 \hat{y}_2 \hat{y}_1 \hat{y}_2+
\hat{y}_2 \hat{y}_2 \hat{y}_1 \hat{y}_2 \hat{y}_2 +
\hat{y}_2 \hat{y}_1 \hat{y}_2 \hat{y}_2 \hat{y}_2+
\hat{y}_1 \hat{y}_2 \hat{y}_2 \hat{y}_2 \hat{y}_2 
)   
  \otimes \left( {\begin{array}{cc}
    0 & 1 \\
   0 & 0 \\
  \end{array} } \right), \,
\nonu \\
&&V^{(\frac{7}{2})+}_{-\frac{5}{2}} = 2(\frac{-i}{4})^{\frac{5}{2}}
\hat{y}_2 \hat{y}_2 \hat{y}_2 \hat{y}_2 \hat{y}_2  \otimes 
  \left( {\begin{array}{cc}
  0 & 1 \\
   0 & 0 \\
  \end{array} } \right),
  \nonu \\
&&V^{(\frac{7}{2})-}_{\frac{5}{2}} = 2(\frac{-i}{4})^{\frac{5}{2}}
\hat{y}_1 \hat{y}_1 \hat{y}_1 \hat{y}_1 \hat{y}_1 \otimes 
  \left( {\begin{array}{cc}
  0 & 0 \\
   1 & 0 \\
  \end{array} } \right), \,
\nonu \\ 
&&V^{(\frac{7}{2})-}_{\frac{3}{2}} = (\frac{-i}{4})^{\frac{5}{2}}
\frac{2}{5}
(\hat{y}_1 \hat{y}_1 \hat{y}_1 \hat{y}_1 \hat{y}_2 +
\hat{y}_1 \hat{y}_1 \hat{y}_1 \hat{y}_2 \hat{y}_1+
\hat{y}_1 \hat{y}_1 \hat{y}_2 \hat{y}_1 \hat{y}_1 +
\hat{y}_1 \hat{y}_2 \hat{y}_1 \hat{y}_1 \hat{y}_1+
\hat{y}_2 \hat{y}_1 \hat{y}_1 \hat{y}_1 \hat{y}_1 
) 
  \otimes \left( {\begin{array}{cc}
   0 & 0 \\
   1 & 0 \\
  \end{array} } \right), 
\nonu \\ 
&&V^{(\frac{7}{2})-}_{\frac{1}{2}} = (\frac{-i}{4})^{\frac{5}{2}}
\frac{2}{10} (\hat{y}_1 \hat{y}_1 \hat{y}_1 \hat{y}_2 \hat{y}_2 +
\hat{y}_1 \hat{y}_1 \hat{y}_2\hat{y}_1 \hat{y}_2 +
\hat{y}_1 \hat{y}_1 \hat{y}_2 \hat{y}_2 \hat{y}_1+
  \hat{y}_1 \hat{y}_2 \hat{y}_1 \hat{y}_1 \hat{y}_2 
   +
   \hat{y}_1 \hat{y}_2 \hat{y}_1\hat{y}_2 \hat{y}_1
   \nonu \\
   && + \hat{y}_1 \hat{y}_2\hat{y}_2 \hat{y}_1 \hat{y}_1
   +\hat{y}_2 \hat{y}_1 \hat{y}_1 \hat{y}_1 \hat{y}_2 +
   \hat{y}_2 \hat{y}_1 \hat{y}_1\hat{y}_2 \hat{y}_1 +
   \hat{y}_2 \hat{y}_1\hat{y}_2 \hat{y}_1 \hat{y}_1+
  \hat{y}_2 \hat{y}_2 \hat{y}_1 \hat{y}_1 \hat{y}_1 
   ) 
  \otimes \left( {\begin{array}{cc}
    0 & 0 \\
   1 & 0 \\
  \end{array} } \right), \,
  \nonu \\
&&V^{(\frac{7}{2})-}_{-\frac{1}{2}} = (\frac{-i}{4})^{\frac{5}{2}}
\frac{2}{10} (\hat{y}_2 \hat{y}_2 \hat{y}_2 \hat{y}_1 \hat{y}_1 +
\hat{y}_2 \hat{y}_2 \hat{y}_1\hat{y}_2 \hat{y}_1 +
\hat{y}_2 \hat{y}_2 \hat{y}_1 \hat{y}_1 \hat{y}_2+
  \hat{y}_2 \hat{y}_1 \hat{y}_2 \hat{y}_2 \hat{y}_1 
   +
   \hat{y}_2 \hat{y}_1 \hat{y}_2\hat{y}_1 \hat{y}_2
   \nonu \\
   && + \hat{y}_2 \hat{y}_1\hat{y}_1 \hat{y}_2 \hat{y}_2
   +\hat{y}_1 \hat{y}_2 \hat{y}_2 \hat{y}_2 \hat{y}_1 +
   \hat{y}_1 \hat{y}_2 \hat{y}_2\hat{y}_1 \hat{y}_2 +
   \hat{y}_1 \hat{y}_2\hat{y}_1 \hat{y}_2 \hat{y}_2+
  \hat{y}_1 \hat{y}_1 \hat{y}_2 \hat{y}_2 \hat{y}_2 
   ) 
  \otimes \left( {\begin{array}{cc}
    0 & 0 \\
   1 & 0 \\
  \end{array} } \right), \,
  \nonu \\
  &&V^{(\frac{7}{2})-}_{-\frac{3}{2}} = (\frac{-i}{4})^{\frac{5}{2}}
\frac{2}{5}
(\hat{y}_2 \hat{y}_2 \hat{y}_2 \hat{y}_2 \hat{y}_1 +
\hat{y}_2 \hat{y}_2 \hat{y}_2 \hat{y}_1 \hat{y}_2+
\hat{y}_2 \hat{y}_2 \hat{y}_1 \hat{y}_2 \hat{y}_2 +
\hat{y}_2 \hat{y}_1 \hat{y}_2 \hat{y}_2 \hat{y}_2+
\hat{y}_1 \hat{y}_2 \hat{y}_2 \hat{y}_2 \hat{y}_2 
)     
  \otimes \left( {\begin{array}{cc}
    0 & 0 \\
   1 & 0 \\
  \end{array} } \right), \,
\nonu \\
&&V^{(\frac{7}{2})-}_{-\frac{5}{2}} = 2(\frac{-i}{4})^{\frac{5}{2}}
\hat{y}_2 \hat{y}_2 \hat{y}_2 \hat{y}_2 \hat{y}_2  \otimes 
  \left( {\begin{array}{cc}
    0 & 0 \\
   1 & 0 \\
  \end{array} } \right),
\nonu \\
&&V^{(4)}_{3} = 2(\frac{-i}{4})^{3}
\hat{y}_1 \hat{y}_1 \hat{y}_1 \hat{y}_1 \hat{y}_1 \hat{y}_1 \otimes 
  \left( {\begin{array}{cc}
  1 & 0 \\
   0 & 1 \\
  \end{array} } \right), \,
\nonu \\ 
&&V^{(4)}_{2} = (\frac{-i}{4})^{3} \frac{2}{6}
(\hat{y}_1 \hat{y}_1 \hat{y}_1 \hat{y}_1 \hat{y}_1  \hat{y}_2+
\hat{y}_1 \hat{y}_1 \hat{y}_1 \hat{y}_1 \hat{y}_2 \hat{y}_1+
\hat{y}_1 \hat{y}_1 \hat{y}_1 \hat{y}_2 \hat{y}_1  \hat{y}_1\nonu \\
&& +
\hat{y}_1 \hat{y}_1 \hat{y}_2 \hat{y}_1 \hat{y}_1 \hat{y}_1+
\hat{y}_1 \hat{y}_2 \hat{y}_1 \hat{y}_1 \hat{y}_1  \hat{y}_1+
\hat{y}_2 \hat{y}_1 \hat{y}_1 \hat{y}_1 \hat{y}_1 \hat{y}_1
)
  \otimes \left( {\begin{array}{cc}
    1 & 0 \\
   0 & 1 \\
  \end{array} } \right), 
\nonu \\ 
&&V^{(4)}_{1} = (\frac{-i}{4})^{3} \frac{2}{15}
(\hat{y}_1 \hat{y}_1 \hat{y}_1 \hat{y}_1 \hat{y}_2  \hat{y}_2
  + \hat{y}_1 \hat{y}_1 \hat{y}_1 \hat{y}_2 \hat{y}_1 \hat{y}_2
  +
  \hat{y}_1 \hat{y}_1 \hat{y}_1 \hat{y}_2 \hat{y}_2 \hat{y}_1+
\hat{y}_1 \hat{y}_1 \hat{y}_2 \hat{y}_1 \hat{y}_1  \hat{y}_2
  + \hat{y}_1 \hat{y}_1 \hat{y}_2 \hat{y}_1 \hat{y}_2 \hat{y}_1
  \nonu \\
  && +
  \hat{y}_1 \hat{y}_1 \hat{y}_2 \hat{y}_2 \hat{y}_1 \hat{y}_1
  +\hat{y}_1 \hat{y}_2 \hat{y}_1 \hat{y}_1 \hat{y}_1  \hat{y}_2
  + \hat{y}_1 \hat{y}_2 \hat{y}_1 \hat{y}_1 \hat{y}_2 \hat{y}_1
  +
  \hat{y}_1 \hat{y}_2 \hat{y}_1 \hat{y}_2 \hat{y}_1 \hat{y}_1
  +\hat{y}_1 \hat{y}_2 \hat{y}_2 \hat{y}_1 \hat{y}_1  \hat{y}_1
  \nonu \\
  && + \hat{y}_2 \hat{y}_1 \hat{y}_1 \hat{y}_1 \hat{y}_1 \hat{y}_2
  +
  \hat{y}_2 \hat{y}_1 \hat{y}_1 \hat{y}_1 \hat{y}_2 \hat{y}_1
  +\hat{y}_2 \hat{y}_1 \hat{y}_1 \hat{y}_2 \hat{y}_1  \hat{y}_1
  + \hat{y}_2 \hat{y}_1 \hat{y}_2 \hat{y}_1 \hat{y}_1 \hat{y}_1
  +
  \hat{y}_2 \hat{y}_2 \hat{y}_1 \hat{y}_1 \hat{y}_1 \hat{y}_1
  ) 
  \otimes \left( {\begin{array}{cc}
    1 & 0 \\
   0 & 1 \\
  \end{array} } \right), 
\nonu \\ 
&&V^{(4)}_{0} = (\frac{-i}{4})^{3}
 \frac{2}{20}
(\hat{y}_1 \hat{y}_1 \hat{y}_1 \hat{y}_2 \hat{y}_2  \hat{y}_2
  + \hat{y}_1 \hat{y}_1 \hat{y}_2 \hat{y}_1 \hat{y}_2 \hat{y}_2
  +
  \hat{y}_1 \hat{y}_1 \hat{y}_2 \hat{y}_2 \hat{y}_1 \hat{y}_2+
\hat{y}_1 \hat{y}_1 \hat{y}_2 \hat{y}_2 \hat{y}_2  \hat{y}_1
 +\hat{y}_1 \hat{y}_2 \hat{y}_1 \hat{y}_1 \hat{y}_2  \hat{y}_2
 \nonu \\
 && + \hat{y}_1 \hat{y}_2 \hat{y}_1 \hat{y}_2 \hat{y}_1 \hat{y}_2
  +
  \hat{y}_1 \hat{y}_2 \hat{y}_1 \hat{y}_2 \hat{y}_2 \hat{y}_1+
\hat{y}_1 \hat{y}_2 \hat{y}_2 \hat{y}_1 \hat{y}_1  \hat{y}_2
 +
  \hat{y}_1 \hat{y}_2 \hat{y}_2 \hat{y}_1 \hat{y}_2 \hat{y}_1+
\hat{y}_1 \hat{y}_2 \hat{y}_2 \hat{y}_2 \hat{y}_1  \hat{y}_1
+ \hat{y}_2 \hat{y}_1 \hat{y}_1 \hat{y}_1 \hat{y}_2 \hat{y}_2
  \nonu \\
  && +
  \hat{y}_2 \hat{y}_1 \hat{y}_1 \hat{y}_2 \hat{y}_1 \hat{y}_2
  +\hat{y}_2 \hat{y}_1 \hat{y}_1 \hat{y}_2 \hat{y}_2  \hat{y}_1
  + \hat{y}_2 \hat{y}_1 \hat{y}_2 \hat{y}_1 \hat{y}_1 \hat{y}_2
  +
  \hat{y}_2 \hat{y}_1 \hat{y}_2 \hat{y}_1 \hat{y}_2 \hat{y}_1
  +\hat{y}_2 \hat{y}_1 \hat{y}_2 \hat{y}_2 \hat{y}_1  \hat{y}_1
  \nonu \\
  && + \hat{y}_2 \hat{y}_2 \hat{y}_1 \hat{y}_1 \hat{y}_1 \hat{y}_2
  +
  \hat{y}_2 \hat{y}_2 \hat{y}_1 \hat{y}_1 \hat{y}_2 \hat{y}_1
  +\hat{y}_2 \hat{y}_2 \hat{y}_1 \hat{y}_2 \hat{y}_1  \hat{y}_1
  + \hat{y}_2 \hat{y}_2 \hat{y}_2 \hat{y}_1 \hat{y}_1 \hat{y}_1
  ) 
  \otimes \left( {\begin{array}{cc}
    1 & 0 \\
   0 & 1 \\
  \end{array} } \right), 
  \nonu \\
&&V^{(4)}_{-1} = (\frac{-i}{4})^{3}
\frac{2}{15}
(\hat{y}_2 \hat{y}_2 \hat{y}_2 \hat{y}_2 \hat{y}_1  \hat{y}_1
  + \hat{y}_2 \hat{y}_2 \hat{y}_2 \hat{y}_1 \hat{y}_2 \hat{y}_1
  +
  \hat{y}_2 \hat{y}_2 \hat{y}_2 \hat{y}_1 \hat{y}_1 \hat{y}_2+
\hat{y}_2 \hat{y}_2 \hat{y}_1 \hat{y}_2 \hat{y}_2  \hat{y}_1
  + \hat{y}_2 \hat{y}_2 \hat{y}_1 \hat{y}_2 \hat{y}_1 \hat{y}_2
  \nonu \\
  && +
  \hat{y}_2 \hat{y}_2 \hat{y}_1 \hat{y}_1 \hat{y}_2 \hat{y}_2
  +\hat{y}_2 \hat{y}_1 \hat{y}_2 \hat{y}_2 \hat{y}_2  \hat{y}_1
  + \hat{y}_2 \hat{y}_1 \hat{y}_2 \hat{y}_2 \hat{y}_1 \hat{y}_2
  +
  \hat{y}_2 \hat{y}_1 \hat{y}_2 \hat{y}_1 \hat{y}_2 \hat{y}_2
  +\hat{y}_2 \hat{y}_1 \hat{y}_1 \hat{y}_2 \hat{y}_2  \hat{y}_2
  \nonu \\
  && + \hat{y}_1 \hat{y}_2 \hat{y}_2 \hat{y}_2 \hat{y}_2 \hat{y}_1
  +
  \hat{y}_1 \hat{y}_2 \hat{y}_2 \hat{y}_2 \hat{y}_1 \hat{y}_2
  +\hat{y}_1 \hat{y}_2 \hat{y}_2 \hat{y}_1 \hat{y}_2  \hat{y}_2
  + \hat{y}_1 \hat{y}_2 \hat{y}_1 \hat{y}_2 \hat{y}_2 \hat{y}_2
  +
  \hat{y}_1 \hat{y}_1 \hat{y}_2 \hat{y}_2 \hat{y}_2 \hat{y}_2
  ) 
  \otimes \left( {\begin{array}{cc}
    1 & 0 \\
   0 & 1 \\
  \end{array} } \right), 
  \nonu \\
  &&V^{(4)}_{-2} = (\frac{-i}{4})^{3} \frac{2}{6}
(\hat{y}_2 \hat{y}_2 \hat{y}_2 \hat{y}_2 \hat{y}_2  \hat{y}_1+
\hat{y}_2 \hat{y}_2 \hat{y}_2 \hat{y}_2 \hat{y}_1 \hat{y}_2+
\hat{y}_2 \hat{y}_2 \hat{y}_2 \hat{y}_1 \hat{y}_2  \hat{y}_2
\nonu \\
&& +
\hat{y}_2 \hat{y}_2 \hat{y}_1 \hat{y}_2 \hat{y}_2 \hat{y}_2+
\hat{y}_2 \hat{y}_1 \hat{y}_2 \hat{y}_2 \hat{y}_2  \hat{y}_2+
\hat{y}_1 \hat{y}_2 \hat{y}_2 \hat{y}_2 \hat{y}_2 \hat{y}_2
)
  \otimes \left( {\begin{array}{cc}
    1 & 0 \\
   0 & 1 \\
  \end{array} } \right), 
  \nonu \\
&&V^{(4)}_{-3} = 2(\frac{-i}{4})^{3}
\hat{y}_2 \hat{y}_2 \hat{y}_2 \hat{y}_2 \hat{y}_2 \hat{y}_2 \otimes 
  \left( {\begin{array}{cc}
  1 & 0 \\
   0 & 1 \\
  \end{array} } \right).
\label{rem}
  \eea
Higher spin generators for (\ref{fields}), in the   extension of $OSp(2|2)$ higher spin algebra,   can be described by the tensor   product between the generators (whose spins and modes   can be fixed by the number of oscillators $\hat{y}_{\al}$)   and the $2 \times 2$   Pauli matrices (plus identity matrix) with appropriate normalization factors in~(\ref{rem}).

\end{document}